\documentclass[aps,pre,onecolumn,superscriptaddress,groupedaddress,nofootinbib]{revtex4-2}

\usepackage{amsmath}  
\usepackage{amssymb}  
\usepackage{graphicx} 
\usepackage{hyperref} 
\usepackage{bm}       
\usepackage[english]{babel} 
\usepackage[normalem]{ulem}

\newcommand{\uE}{\mathrm{E}}
\newcommand{\blue}[1]{\textcolor{blue}{#1}}

\hypersetup{
    colorlinks=true,
    linkcolor=blue,
    citecolor=blue,
    urlcolor=blue
}

\begin{document}

\title{Time Evolution of Heat Conduction in a Generalized Model of Brownian Motion}

\author{T. Koide}
\email{tomoikoide@gmail.com}
\affiliation{Instituto de F\'{\i}sica, Universidade Federal do Rio de Janeiro, 21941-972, Rio de Janeiro, RJ, Brazil}

\author{F. Nicacio}
\email{nicacio@if.ufrj.br}
\affiliation{Instituto de F\'{\i}sica, Universidade Federal do Rio de Janeiro, 21941-972, Rio de Janeiro, RJ, Brazil}

\date{\today}

\begin{abstract}
We investigate the properties of heat conduction in a network of harmonic oscillators interacting
with heat baths, described by a generalized model of Brownian motion. This model includes noise and
dissipation terms in both the momentum and position equations. 
This generalization is motivated by the
requirement of consistency with the Gorini-Kossakowski-Sudarshan-Lindblad (GKSL) equation. 
Because standard definitions of heat current based on velocity become mathematically inconsistent in this
framework, 
we derive an analytical expression for the steady-state heat flow based on an extended
framework of stochastic energetics. 
We confirm that Fourier's law (linear thermal response) is
satisfied and that the model naturally captures microscopic thermal boundary resistance, 
analogous to Kapitza resistance. 
This demonstrates that our generalized model functions as a valid phenomenological framework for simulating
non-equilibrium processes, 
marking a crucial step toward a unified formulation of stochastic and quantum thermodynamics. 
Furthermore, we analyze the time evolution of heat conduction 
by numerically solving the corresponding differential equations
for the correlation functions.
Unlike standard Brownian motion, the generalized model generates continuous 
and nowhere differentiable trajectories for both momentum and position (as is characteristic of
overdamped dynamics). 
Finally, we show that the heat current exhibits characteristic transient behavior when the inter-particle interaction is switched on.
Specifically, an instantaneous heat flow emerges, whose direction is strictly governed by whether
the interaction is attractive or repulsive, significantly differing from the predictions of the
standard model.
\end{abstract}

\maketitle


\section{Introduction}
\label{sec:intro}

The derivation of macroscopic thermal conduction laws, such as Fourier's law, from microscopic dynamics remains a cornerstone problem in non-equilibrium statistical mechanics. 
To study thermal transport from a microscopic perspective, low-dimensional lattice models coupled to heat reservoirs via stochastic differential equations have been employed \cite{RLL1967,Jackson1978,Lepri2003, Dhar2008,Bonetto2000,Sekimoto2010}. 
In standard models of Brownian motion, the environment is assumed to affect only 
the momentum differential equation of the Brownian particles. 
Consequently, fluctuation and dissipation are added exclusively to the momentum equation, while the kinematic relation for position and momentum maintains the standard linear form ($\dot{q} = p/m$).

This standard framework has proven to be physically insufficient for providing a proper classical limit for open quantum systems.
Recently, the present authors demonstrated that deriving a physically consistent quantum master equation from classical stochastic models via canonical quantization requires a generalized framework \cite{KN2024,KN2024_CPTP,KN2025}.
Specifically, to reproduce the Gorini-Kossakowski-Sudarshan-Lindblad (GKSL) equation, which is a
linear quantum master equation satisfying the Completely Positive and Trace-Preserving (CPTP)
condition \cite{Breuerbook}, the underlying classical Brownian motion must incorporate fluctuation
and dissipation effects into both the momentum and position coordinates.
The same generalized model of Brownian motion has also been explored by other
authors \cite{Giordano}, who reiterate our argument.
While introducing a noise term into the position equation alters the standard proportional relation
between velocity and momentum, such a deviation is a natural feature in the presence of
velocity-dependent forces, analogous to the Lorentz force in electromagnetism.
Since thermal relaxation is typically induced by a velocity-dependent force,
it is thus natural to expect a modification of the standard proportional
relation in our model.
%
%
Furthermore, as a direct mathematical consequence of this generalization, the position trajectories of the Brownian particles in our model become continuous but nowhere differentiable. This is because while conventional Brownian models in phase space yield smooth position trajectories through the time integration of stochastic momentum, our approach introduces fluctuation effects directly into both the position and momentum equations. As a result, a fundamental question arises: can a model with such drastically altered microscopic trajectories still reproduce macroscopic non-equilibrium behavior, such as Fourier's law of heat conduction?

Motivated by
%
%
this quantum-classical correspondence, we investigate the thermal conduction properties of a coupled
harmonic oscillator network described by the generalized model of Brownian motion. 
While our previous work has shown 
that this framework satisfies the first and second laws of stochastic thermodynamics, its
phenomenological validity for non-equilibrium heat transport must be rigorously tested.
In this paper, we focus on two primary aspects. 
First, we analyze the non-equilibrium steady state to analytically verify the emergence of a Fourier-like linear thermal response. 
We also examine whether the model can microscopically capture realistic interfacial transport phenomena, such as the temperature discontinuity known as thermal boundary resistance (akin to Kapitza resistance) \cite{Swartz1989,Giri2020}. 
Second, we explore the transient dynamics of the heat currents when the inter-particle interaction is switched on.
We specifically investigate how the fluctuations on positions and the nature of the interaction (attractive versus repulsive) govern the initial energy flow and instantaneous
heat current jumps.

Before proceeding, a brief remark on terminology is necessary. In statistical physics, 
an integro-differential equation with a memory kernel is often referred to as a generalized Langevin equation, see for instance \cite{book:evang}. 
We explicitly distinguish our approach from such models. The generalized model of Brownian motion considered in this paper extends the standard framework by introducing fluctuation and dissipation directly into both position and momentum equations, rather than relying on memory-based generalizations.

The paper is organized as follows. 
In Sec.~\ref{sec:model}, we introduce the network model and the generalized stochastic differential equations. 
In Sec.~\ref{sec:Heat_and_Thermodynamics}, we define heat and work based on stochastic thermodynamics. 
Section~\ref{sec:steady_state} is devoted to the steady-state analysis and the verification of
Fourier's law (linear thermal response).
In Sec.~\ref{sec:transient}, we discuss the time evolution of heat currents and the effect of
the positional noise parameter. `Finally, we summarize our results in Sec.~\ref{sec:conclusion}.

\section{Network Model}
\label{sec:model}

\subsection{Generalized Brownian Motion}

We consider a system of $N$ particles interacting with heat reservoirs.
For simplicity, we assume a one-dimensional system.
Following the generalized model of Brownian motion proposed in Ref.~\cite{KN2024}, the stochastic time evolution of the system is described by the following set of stochastic differential equations:
\begin{align}
    d\tilde{q}_{i,t} &= \left( \frac{\partial H}{\partial \tilde{p}_{i,t}} - \gamma_{q_i}
\frac{\partial H}{\partial \tilde{q}_{i,t}} \right) dt + \sqrt{\frac{2\gamma_{q_i}}{\beta_i}}
d\tilde{B}_{{q}_i,t} \, , \label{eq:gen_dq} \\
    d\tilde{p}_{i,t} &= \left( -\frac{\partial H}{\partial \tilde{q}_{i,t}} - \gamma_{p_i}
\frac{\partial H}{\partial \tilde{p}_{i,t}} \right) dt + \sqrt{\frac{2\gamma_{p_i}}{\beta_i}}
d\tilde{B}_{{p}_i,t} \, , \label{eq:gen_dp}
\end{align}
where $\tilde{q}_{i,t}$ and $\tilde{p}_{i,t}$ represent the stochastic position and momentum of the
$i$-th particle at time $t$, respectively.
Hereafter, the tilde symbol ``$\tilde{\phantom{a}}$'' is used to explicitly denote stochastic
variables.
The function $H(\{\tilde{q}_t\}, \{\tilde{p}_t\})$ represents our system Hamiltonian, which can take
an arbitrary form.
The parameters $\gamma_{q_i}$ and $\gamma_{p_i}$ are non-negative real constants representing the
dissipation coefficients for the position and momentum coordinates, respectively.
The system interacts with heat baths at inverse temperatures $\beta_i := 1/(k_B T_i)$, with $k_B$
being the Boltzmann constant.
The noise terms $d\tilde{B}_{{q}_i,t}$ and $d\tilde{B}_{{p}_i,t}$ are independent Wiener processes
satisfying the following correlation properties:
\begin{align}
    \uE[d\tilde{B}_{\alpha,i,t}] &= 0 \, , \label{eq:noise_mean} \\
    \uE[d\tilde{B}_{\alpha,i,t} \, \, d\tilde{B}_{\delta,j,t'}] &= \delta_{\alpha,\delta}
\delta_{i,j} \delta_{t,t'} dt \, , \label{eq:noise_corr}
\end{align}
where $\alpha, \delta \in \{q, p\}$ and $\uE[\, \cdot \,]$ denotes the ensemble average.
Standard Brownian motion corresponds to the limit $\gamma_{q_i} \to 0$.
The noise strengths are determined by the fluctuation-dissipation theorem (Einstein relation) to
ensure thermodynamic consistency.
As shown in Refs.~\cite{KN2024,KN2024_CPTP,KN2025}, this generalized model satisfies the laws
analogous to those of thermodynamics.
Heat, which will be given in Sec.\ \ref{sec:Heat_and_Thermodynamics}, is defined as the work done by
the stochastic forces from the heat baths.
Under this definition, the first law of thermodynamics is satisfied as an energy balance equation
for each stochastic event.
Furthermore, using the Shannon entropy defined for the phase space distribution function, our generalized model of Brownian motion satisfies the second law of thermodynamics in the form of the Clausius inequality.

\subsection{Two Harmonic Oscillators} 

Traditionally, to investigate lattice heat conduction from microscopic dynamics, models consisting
of $N$ linearly interacting harmonic oscillators have been extensively studied
\cite{RLL1967,Jackson1978,Lepri2003, Dhar2008,Bonetto2000}.
In contrast, in this work, we begin our analysis with the minimal network model of $N=2$ particles
to rigorously and analytically derive the emergence of Fourier's law (linear thermal response)
within the generalized model of Brownian motion.
The system Hamiltonian is given by 
\begin{equation}
    H (\{ \tilde{q}_{i,t}, \tilde{p}_{i,t} \}) = \sum_{i=1}^{2} \left[
\frac{\tilde{p}_{i,t}^2}{2m_i} + \frac{K_i}{2}(\tilde{q}_{i,t} - x_{(i)})^2 \right] + V_\text{int}
\, ,
    \label{eq:hamiltonian}
\end{equation}
where $m_i$ and $K_i$ are the mass and spring constant of the $i$-th oscillator, and $x_{(i)}$
denotes the center of oscillation for each particle.
The interaction potential $V_\text{int}$ is defined by
\begin{equation}
    V_\text{int} = \frac{K_\text{int}}{2}(\tilde{q}_{1,t} - \tilde{q}_{2,t} - l_0)^2 \, ,
    \label{eq:interaction}
\end{equation}
where $K_\text{int}$ is the coupling constant and $l_0 = x_{(1)}-x_{(2)}$ is a natural length.

Substituting Eq.~\eqref{eq:hamiltonian} into Eqs.~\eqref{eq:gen_dq} and \eqref{eq:gen_dp},
we obtain the specific equations of motion for our system:
\begin{align}
    d\tilde{q}_{i,t} &= \left[ \frac{\tilde{p}_{i,t}}{m_i} - \gamma_{q_i} \left( K_i(\tilde{q}_{i,t}
- x_{(i)}) + \frac{\partial V_\text{int}}{\partial \tilde{q}_{i,t}} \right) \right] dt +
\sqrt{\frac{2\gamma_{q_i}}{\beta_i}} d\tilde{B}_{{q}_i,t} \, , \label{eq:sde_q} \\
    d\tilde{p}_{i,t} &= \left[ - \left( K_i(\tilde{q}_{i,t} - x_{(i)}) + \frac{\partial
V_\text{int}}{\partial \tilde{q}_{i,t}} \right) - \gamma_{p_i} \frac{\tilde{p}_{i,t}}{m_i} \right]
dt + \sqrt{\frac{2\gamma_{p_i}}{\beta_i}} d\tilde{B}_{{p}_i,t} \, , \label{eq:sde_p}
\end{align}
where the partial derivatives of the interaction potential are given by 
\begin{align}
\frac{\partial V_\text{int}}{\partial \tilde{q}_{1,t}} &= K_\text{int}(\tilde{q}_{1,t} -
\tilde{q}_{2,t}
- l_0) \, ,\\
\frac{\partial V_\text{int}}{\partial \tilde{q}_{2,t}} &= -K_\text{int}(\tilde{q}_{1,t} -
\tilde{q}_{2,t}
- l_0) \, .
\end{align}

The Wiener processes satisfy the standard correlation properties defined in
Eqs.~\eqref{eq:noise_mean} and \eqref{eq:noise_corr}.
We assume that the two oscillators are connected to separate heat baths with temperatures $T_1$ and
$T_2$, respectively, and we will analytically demonstrate that heat conduction is induced by the temperature difference
$\Delta T = T_1 - T_2$.

\section{Heat and Thermodynamics} \label{sec:Heat_and_Thermodynamics}

Based on stochastic thermodynamics \cite{Sekimoto2010,seifert2012}, 
let $d\tilde{Q}_{i,t}$ denote the heat absorbed by the system from the $i$-th heat bath during an
infinitesimal time interval $dt$.
Following the stochastic energetics approach, we define the heat as the work done by the forces
(fluctuation and dissipation) exerted by the heat baths, \textit{i.e.}, the work of the external forces acting on the system.
Unlike standard stochastic thermodynamics, 
our generalized model considers additional forces acting on the position equations, 
namely the terms $- \gamma_{q_i}
(\partial H/ \partial \tilde{q}_{i,t}) dt$ and $\sqrt{2\gamma_{q_i}/\beta_i}
d\tilde{B}_{{q}_i,t}$ 
in Eq.~(\ref{eq:sde_q}).
Therefore, the standard definition of heat \cite{Sekimoto2010} must be extended to include 
these contributions from the position variable, which were proposed in Ref.~\cite{KN2024}:
\begin{align}
    d\tilde{Q}_{i,t} &:= \left( -\gamma_{p_i} \frac{\partial H}{\partial \tilde{p}_{i,t}} dt +
\sqrt{\frac{2\gamma_{p_i}}{\beta_i}} d\tilde{B}_{{p}_i,t} \right) \circ \frac{\partial H}{\partial
\tilde{p}_{i,t}} \nonumber \\
    &\quad + \left( -\gamma_{q_i} \frac{\partial H}{\partial \tilde{q}_{i,t}} dt +
\sqrt{\frac{2\gamma_{q_i}}{\beta_i}} d\tilde{B}_{{q}_i,t} \right) \circ \frac{\partial H}{\partial
\tilde{q}_{i,t}} \, ,
    \label{eq:heat_def}
\end{align}
where the symbol ``$\circ$" denotes the Stratonovich definition of the product of stochastic
variables \cite{book:gardiner}.
The first term in the above equation corresponds to the work done by the random force and
dissipation associated with $ d\tilde{p}_{i,t}$, which is the standard definition of heat in stochastic energetics,
while the second is the new contribution arising specifically from the noise and dissipation
associated with $ d\tilde{q}_{i,t}$.
This is the natural generalization of the mesoscopic heat proposed by Sekimoto \cite{Sekimoto2010}, 
which is reproduced in the vanishing limit of $\gamma_{q_i}$.

Since the present system lacks external forces that explicitly deform the Hamiltonian, the first law
of thermodynamics relates the change in the system's Hamiltonian to the total heat absorbed.
Thus, for the case of $N=2$, this reads:
\begin{equation}
    d H (\{ \tilde{q}_{i,t}, \tilde{p}_{i,t} \}) = \sum_{i=1}^{2} d\tilde{Q}_{i,t} \, .
    \label{eq:first_law}
\end{equation}
Let us calculate the expectation value of this first law. 
By taking the ensemble average of Eq.~\eqref{eq:heat_def} and using It\^o's lemma,  
the expectation value of the heat current from the $i$-th bath is given by
\begin{align}
    J_i(t) := \left\lceil \frac{d\tilde{Q}_{i,t}}{dt} \right\rfloor&= -\frac{\gamma_{p_i}}{m_i^2}
\left\lceil \tilde{p}_{i,t}^2 \right\rfloor \nonumber \\
    &\quad - \gamma_{q_i} \left\lceil \left( K_i(\tilde{q}_{i,t} - x_{(i)}) + \frac{\partial
V_\text{int}}{\partial \tilde{q}_{i,t}} \right)^2 \right\rfloor \nonumber \\
    &\quad + \frac{\gamma_{p_i}}{\beta_i m_i} + \frac{\gamma_{q_i}}{\beta_i} \left( K_i +
K_\text{int}
\right) \, ,
    \label{eq:mean_heat_current}
\end{align}
where we introduced a symbol representing the average over the initial distribution of the phase
space variables and event ensemble of the Wiener process:
\begin{equation}
\lceil \tilde{A}  \rfloor :=
\int d q_{1,0} d q_{2,0} \int d p_{1,0} dp_{2,0}
\rho_0 (q_{1,0},q_{2,0},p_{1,0},p_{2,0}) \uE \left[ \tilde{A}(\{ \tilde{q}_{i,t}, \tilde{p}_{i,t}
\}) \right] \, , \label{eq:average}
\end{equation}    
where $\rho_0 (q_{1,0},q_{2,0},p_{1,0},p_{2,0})$ is an initial phase space distribution normalized
to unity.
Here we have used the system Hamiltonian defined by Eqs.\ (\ref{eq:hamiltonian}) and
(\ref{eq:interaction}).
From this expression, we can confirm the consistency with the first law of thermodynamics in terms
of expectation values:
\begin{equation}
    \frac{d}{dt} \left\lceil H\right\rfloor = \sum_{i=1}^{2} J_i(t).
    \label{eq:first_law_avg}
\end{equation}
In the steady state ($t \to \infty$), the total energy of the system becomes constant 
($d \left\lceil H\right\rfloor/dt = 0$), implying the balance of heat currents:
\begin{equation}
J_1^\text{ss} = -J_2^\text{ss} \, ,
\end{equation}
meaning the heat flowing out of the hot reservoir equals the heat flowing into the cold reservoir.

\section{Fourier's Law in Heat Conduction}
\label{sec:steady_state}

To investigate whether Fourier's law (linear thermal response) holds in the steady state, we
consider a simplified model.
We assume that the two particles have identical masses and spring constants, and we set the center
of oscillation and natural length to zero:
\begin{align}
\begin{split} \label{eq:simplifications}
&m_1 = m_2 = m \, , \\
&K_1 = K_2 = K \, , \\
&x_{(i)} = 0 \, ,\\
&l_0 = 0 \, .
\end{split}
\end{align}

To simplify the analysis, we introduce the position normal mode coordinates
$\tilde{R}_{(i)t}$ by
\begin{equation}
    \begin{pmatrix} \tilde{R}_{(1)t} \\ \tilde{R}_{(2)t} \end{pmatrix} = \frac{1}{\sqrt{2}}
\begin{pmatrix} \tilde{q}_{1,t} - \tilde{q}_{2,t} \\ \tilde{q}_{1,t} + \tilde{q}_{2,t}
\end{pmatrix} \, , \label{eq:coordinate_trans}
\end{equation}
which transform the Hamiltonian (\ref{eq:hamiltonian}) into the one of two decopled harmonic
oscillators,
characterized by the effective spring constants $\lambda_{\pm}$ given by
\begin{equation}\label{eq:lambda}
    \lambda_{\pm} = K + K_\text{int} \pm K_\text{int} \, ,
\end{equation}
where $\lambda_+ = K + 2K_\text{int}$ and $\lambda_- = K$.
To ensure that the normal modes are oscillatory, these effective constants must be strictly positive:
\begin{equation}
 \lambda_\pm > 0 \, . \label{eqn:positive_lambda}
\end{equation}

We now introduce adimensional quantities to simplify the equations.
We define the characteristic scales $t_0, q_0, p_0$ which have dimensions of time, length and
momentum, respectively.
Then we can define the adimensional variables as follows:
\begin{align} \label{eq:adim_factors}
\begin{split}
\overline{t}  &= \frac{t}{t_0}\, , \\
\overline{p}_{(i)\overline{t}} &= \frac{\tilde{p}_{i,t} }{p_0 } \, , \\
\overline{R}_{(i)\overline{t}} &= \frac{\tilde{R}_{(i)t} }{q_0 }\, .
\end{split}
\end{align}

Using (\ref{eq:average}), we can define the (adimensional) second moments:
\begin{align}
\begin{split}
\Delta_{p_i p_j} &= \frac{\left\lceil \tilde{p}_{i,t} \tilde{p}_{j,t} \right\rfloor}{p_0^2} \, ,
\\
\Delta_{p_i R_j} &= \frac{\left\lceil \tilde{p}_{i,t} \tilde{R}_{(j)t} \right\rfloor}{p_0 q_0} \, ,
\\
\Delta_{R_i R_j} &= \frac{\left\lceil \tilde{R}_{(i)t} \tilde{R}_{(j)t} \right\rfloor}{q_0^2 } \, ,
\end{split}\label{eq:delta_elem}
\end{align}
which can be organized in the vector $\boldsymbol{\Delta}$ consisting of the
following components:
\begin{equation}
    \boldsymbol{\Delta} = (\Delta_{p_1 p_1}, \Delta_{p_2 p_2}, \Delta_{p_1 p_2}, \Delta_{p_1 R_1},
\Delta_{p_1 R_2}, \Delta_{p_2 R_1}, \Delta_{p_2 R_2}, \Delta_{R_1 R_1}, \Delta_{R_2 R_2},
\Delta_{R_1 R_2})^T.
    \label{eq:delta_def}
\end{equation}

The time evolutions of these components can be derived using
It\^o's lemma \cite{book:gardiner}. 
For instance,
let us consider the element $\Delta_{p_1 R_2}$.
According to Eq.~(\ref{eq:delta_elem}), its variation is determined by
\begin{equation}
\begin{aligned}
d(\tilde{p}_{1,t}\tilde{R}_{(2)t}) =
& \,\, \tilde{R}_{(2)t} d\tilde{p}_{1,t} + \tilde{p}_{1,t} d\tilde{R}_{(2)t} \\
= &- \tilde{R}_{(2)t}
\left\{ K_i \tilde{q}_{1,t}  +  K_\text{int}(\tilde{q}_{1,t}-\tilde{q}_{2,t}) \right\}
+
\tilde{R}_{(2)t}
\left(
- \gamma_{p_1}\frac{\tilde{p}_{1,t}}{m}\,dt
+ \sqrt{\frac{2\gamma_{p_1}}{\beta_1}}\, d\tilde{B}_{p_i , t}
\right) \\
&+ \frac{\tilde{p}_{1,t}}{\sqrt{2}}
\left\{
\frac{\tilde{p}_{1,t} + \tilde{p}_{2,t}}{m}\,dt
- \frac{\lambda_+ (\gamma_{q_1} - \gamma_{q_2})}{\sqrt{2}} \tilde{R}_{(1)t}\,dt
- \frac{\lambda_- (\gamma_{q_1} + \gamma_{q_2})}{\sqrt{2}} \tilde{R}_{(2)t}\,dt \right\} \\
&+ \frac{\tilde{p}_{1,t}}{\sqrt{2}}
\left\{\sqrt{\frac{2\gamma_{q_1}}{\beta_1}}\, d\tilde{B}_{q_1,t}
+ \sqrt{\frac{2\gamma_{q_2}}{\beta_2}}\, d\tilde{B}_{q_2 , t}
\right\} \, ,
\end{aligned}
\end{equation}
where we used $d\tilde{p}_{1,t}$ from Eq.~(\ref{eq:sde_p}). 
According to Eq.~(\ref{eq:coordinate_trans}), the variation $d\tilde{R}_{(2)t}$ is given by $\sqrt{2} \, d\tilde{R}_{(2)t} = d\tilde{q}_{1,t} + d\tilde{q}_{2,t}$, where $d\tilde{q}_{i,t}$ is taken from Eq.~(\ref{eq:sde_q}). 
The terms were then rearranged by using the definition of $\lambda_\pm$ in Eq.~(\ref{eq:lambda}).
Note that second-order terms such as $d\tilde{p}_{1,t} d\tilde{R}_{(2)t}$ do not contribute to this stochastic calculus.
By taking the expectation value,using 
the fact that the Wiener process has zero mean due to Eq.~(\ref{eq:noise_mean}), and rearranging the terms, we find
\begin{equation}
\begin{aligned}
\frac{d}{d\overline{t}}\Delta_{p_1 R_2} = \frac{1}{d\overline{t}}\left\lceil
\frac{d(\tilde{p}_{1,t}\tilde{R}_{(2)t})}{q_0p_0} \right\rfloor
= & - \frac{\lambda_+}{\sqrt{2}} \frac{q_0 t_0}{p_0} \Delta_{R_1 R_2}
- \frac{\lambda_-}{\sqrt{2}} \frac{q_0 t_0}{p_0} \Delta_{R_2 R_2}
- \frac{\gamma_{p_1} t_0}{m} \Delta_{p_1 R_2}
+ \frac{1}{\sqrt{2m}} \frac{p_0 t_0}{q_0} \Delta_{p_1 p_1} \\
& + \frac{1}{\sqrt{2m}} \frac{p_0 t_0}{q_0} \Delta_{p_1 p_2}
- \frac{\lambda_+ (\gamma_{q_1} - \gamma_{q_2})}{2} t_0 \Delta_{p_1 R_1}
- \frac{\lambda_- (\gamma_{q_1} + \gamma_{q_2})}{2} t_0 \Delta_{p_1 R_2} \, ,
\end{aligned}
\end{equation}
where we used the definitions in Eq.~(\ref{eq:delta_elem}).
Following the same procedure for all the other elements of ${\bf\Delta}$ in Eq.~(\ref{eq:delta_def}), we obtain the system of differential equations for these elements as
\begin{equation}
    \frac{d}{d\overline{t}} \boldsymbol{\Delta} = \mathbf{M} \boldsymbol{\Delta} + \mathbf{C} \, .
    \label{eq:matrix_evolution}
\end{equation}
where
\begin{equation}
    \mathbf{C} = (
    2\overline{\gamma}_{p_1}/\overline{\beta}_1,
    2\overline{\gamma}_{p_2}/\overline{\beta}_2, 
    0,
    0,
    0,
    0,
    0,
    \overline{\gamma}_{q_1}/\overline{\beta}_1 + \overline{\gamma}_{q_2}/\overline{\beta}_2,
    \overline{\gamma}_{q_1}/\overline{\beta}_1 + \overline{\gamma}_{q_2}/\overline{\beta}_2,
    \overline{\gamma}_{q_1}/\overline{\beta}_1 - \overline{\gamma}_{q_2}/\overline{\beta}_2
    )^T \, ,
\end{equation}
is a constant row vector defined by the adimensional quantities:
\begin{align} \label{eq:adimensional}
\begin{split}
\overline{\beta}_i  &= \frac{p_0^2 \beta_i}{m}\, , \\
\overline{\gamma}_{p_i} &= \frac{t_0 \gamma_{p_i}}{m} \, ,\\
\overline{\gamma}_{q_i} &= \frac{p_0 \gamma_{q_i}}{q_0 } \, .
\end{split}
\end{align}
At this point, we should note that the coefficient matrix $\mathbf{M}$ is a $10 \times 10$ matrix defined as follows:
\begin{equation}
\mathbf{M} =
\begin{scriptsize}
\left(
\begin{array}{cccccccccc}
 -2 \overline{\gamma}_{p_1} & 0 & 0 & -\sqrt{2} \overline{\lambda}_+ & -\sqrt{2}
\overline{\lambda}_- & 0 & 0 & 0 & 0 & 0 \\
 0 & -2 \overline{\gamma}_{p_2} & 0 & 0 & 0 & \sqrt{2} \overline{\lambda}_+ & -\sqrt{2}
\overline{\lambda}_- & 0 & 0 & 0 \\
 0 & 0 & -(\overline{\gamma}_{p_1}+\overline{\gamma}_{p_2}) & \frac{\overline{\lambda}_+}{\sqrt{2}}
& -\frac{\overline{\lambda}_-}{\sqrt{2}} & -\frac{\overline{\lambda}_+}{\sqrt{2}} &
-\frac{\overline{\lambda}_-}{\sqrt{2}} & 0 & 0 & 0 \\
 \frac{1}{\sqrt{2}} & 0 & -\frac{1}{\sqrt{2}} & M_{4,4} & M_{4,5} & 0 & 0 &
-\frac{\overline{\lambda}_+}{\sqrt{2}} & 0 & -\frac{\overline{\lambda}_-}{\sqrt{2}} \\
 \frac{1}{\sqrt{2}} & 0 & \frac{1}{\sqrt{2}} & M_{5,4} & M_{5,5} & 0 & 0 & 0 &
-\frac{\overline{\lambda}_-}{\sqrt{2}} & -\frac{\overline{\lambda}_+}{\sqrt{2}} \\
 0 & -\frac{1}{\sqrt{2}} & \frac{1}{\sqrt{2}} & 0 & 0 & M_{6,6} & M_{6,7} &
\frac{\overline{\lambda}_+}{\sqrt{2}} & 0 & -\frac{\overline{\lambda}_-}{\sqrt{2}} \\
 0 & \frac{1}{\sqrt{2}} & \frac{1}{\sqrt{2}} & 0 & 0 & M_{7,6} & M_{7,7} & 0 &
-\frac{\overline{\lambda}_-}{\sqrt{2}} & \frac{\overline{\lambda}_+}{\sqrt{2}} \\
 0 & 0 & 0 & \sqrt{2} & 0 & -\sqrt{2} & 0 & M_{8,8} & 0 & M_{8,10} \\
 0 & 0 & 0 & 0 & \sqrt{2} & 0 & \sqrt{2} & 0 & M_{9,9} & M_{9,10} \\
 0 & 0 & 0 & \frac{1}{\sqrt{2}} & \frac{1}{\sqrt{2}} & \frac{1}{\sqrt{2}} & -\frac{1}{\sqrt{2}} &
M_{10,8} & M_{10,9} & M_{10,10} \\
\end{array}
\right) \, ,
\end{scriptsize}
\end{equation}
where 
\begin{equation}
\overline{\lambda}_{\pm} = \lambda_{\pm} \frac{q_0 t_0}{p_0} \, ,
\label{eq:adim_lambda}
\end{equation}
and, for notational brevity, some elements are expressed using the parameters $M_{i,j}$,
which are defined by
\begin{align}
\begin{split}
M_{4,4} &= -\overline{\gamma}_{p_1} -
\frac{\overline{\lambda}_+(\overline{\gamma}_{q_1}+\overline{\gamma}_{q_2})}{2} \, , \\
M_{4,5} &= -\frac{\overline{\lambda}_-(\overline{\gamma}_{q_1}-\overline{\gamma}_{q_2})}{2}\, ,
\\
M_{5,4} &= -\frac{\overline{\lambda}_+(\overline{\gamma}_{q_1}-\overline{\gamma}_{q_2})}{2} \,
,\\
M_{5,5} &= -\overline{\gamma}_{p_1} -
\frac{\overline{\lambda}_-(\overline{\gamma}_{q_1}+\overline{\gamma}_{q_2})}{2} \, , \\
M_{6,6} &= -\overline{\gamma}_{p_2} -
\frac{\overline{\lambda}_+(\overline{\gamma}_{q_1}+\overline{\gamma}_{q_2})}{2} \, , \\
M_{6,7} &= -\frac{\overline{\lambda}_-(\overline{\gamma}_{q_1}-\overline{\gamma}_{q_2})}{2} \,
, \\
M_{7,6} &= -\frac{\overline{\lambda}_+(\overline{\gamma}_{q_1}-\overline{\gamma}_{q_2})}{2} \,
, \\
M_{7,7} &= -\overline{\gamma}_{p_2} -
\frac{\overline{\lambda}_-(\overline{\gamma}_{q_1}+\overline{\gamma}_{q_2})}{2} \, , \\
M_{8,8} &= -\overline{\lambda}_+(\overline{\gamma}_{q_1}+\overline{\gamma}_{q_2}) \, , \\
M_{8,10} &= -\overline{\lambda}_-(\overline{\gamma}_{q_1}-\overline{\gamma}_{q_2}) \, , \\
M_{9,9} &= -\overline{\lambda}_-(\overline{\gamma}_{q_1}+\overline{\gamma}_{q_2}) \, , \\
M_{9,10} &= -\overline{\lambda}_+(\overline{\gamma}_{q_1}-\overline{\gamma}_{q_2}) \, , \\
M_{10,8} &= M_{10,9} =
-\frac{\overline{\lambda}_-(\overline{\gamma}_{q_1}-\overline{\gamma}_{q_2})}{2} \, , \\
M_{10,10} &= -\frac{(\overline{\lambda}_+ +
\overline{\lambda}_-)(\overline{\gamma}_{q_1}+\overline{\gamma}_{q_2})}{2} \, .
\end{split}
\end{align}

\subsection{Simplified Limit and Decoupling}

In order to derive Fourier's law (linear thermal response),
we proceed with the analytical treatment of Eq.~(\ref{eq:matrix_evolution}) by assuming symmetric dissipation coefficients in Eqs.~(\ref{eq:gen_dq}) and
(\ref{eq:gen_dp}):
\begin{align}\label{eq:uniform_couplings}
    \gamma_{p_1} &= \gamma_{p_2} := \gamma_p \, ,\\
    \gamma_{q_1} &= \gamma_{q_2} := \gamma_q \, .
\end{align}
In standard Brownian motion, ${\gamma}_q$ vanishes for a finite ${\gamma}_p$.
We thus introduce a parameter $\delta \ge -1$ to characterize ${\gamma}_q$ in terms of
${\gamma}_p \ge 0$.
In the adimensional form,
see Eq.'(\ref{eq:adimensional}), we parametrize $\bar{\gamma}_q$ as
\begin{equation}
    \overline{\gamma}_q = \overline{\gamma}_p (1 + \delta) \ge 0  \, . \label{eqn:gamma_delta}
\end{equation}
Beside the scalling in Eqs.~(\ref{eq:adimensional}),
we further choose the time scale $t_0$ and length scale ratio $q_0/p_0$ as
\begin{align}
t_0 &= \sqrt{\frac{m}{K}} \, ,\\
\frac{q_0 t_0}{p_0} &= \frac{1}{K} \, .
\end{align}
With these choices and their applications to Eq.~(\ref{eq:adim_lambda}),
the adimensional eigenvalues $\overline{\lambda}_{\pm}$ are further simplified:
\begin{equation}
    \overline{\lambda}_{\pm} = 1 + \chi \pm \chi \, ,
\end{equation}
where
\begin{equation}\label{eq:chi}
   \chi = \frac{K_\text{int}}{K} \, ,
\end{equation}
is the ration between the individual and the interaction spring constants of the system,
see Eqs.~(\ref{eq:hamiltonian}), (\ref{eq:interaction}), and (\ref{eq:simplifications}).
Specifically, $\overline{\lambda}_+ = 1 + 2\chi$ and $\overline{\lambda}_- = 1$.
It should be emphasized that $\chi$ can take negative values, 
but, due to the condition in Eq.~(\ref{eqn:positive_lambda}),
restricted to the following domain:
\begin{equation}
    -0.5 < \chi < \infty \, . \label{eqn:para_chi}
\end{equation}

The heat current difference $J_1-J_2$, 
using (\ref{eq:mean_heat_current}) and the elements in Eq.(\ref{eq:delta_elem}), is then written as
\begin{equation}
J_1-J_2 = \frac{p_0^2 \overline{\gamma}_{p} }{t_0 m} \left[
(\Delta_{p_2 p_2} - \Delta_{p_1 p_1})
- 2 (1 + \delta)(1 + 2\chi) \Delta_{R_1 R_2}
+ \{ 1+ (1+\delta)  (1+\chi) \} \Delta \mathcal{T}
\right] \, , \label{eq:diff_currents}
\end{equation}
where we define the temperature difference of the heat baths:
\begin{align}\label{eq:dif_temp}
\Delta \mathcal{T} &:= \frac{1}{\overline{\beta}_1} - \frac{1}{\overline{\beta}_2} \, .
\end{align}
Note that the relevant quantities for the currents difference $J_1-J_2$ are the correlations $\Delta_{p_1 p_1} - \Delta_{p_2 p_2}$ and $\Delta_{R_1 R_2}$. 
We will show that these correlations are linear functions of $\Delta \mathcal{T}$.

From the general matrix equation \eqref{eq:matrix_evolution}, we can extract the time evolution
equations for the specific linear combinations of correlation functions:
\begin{align}
    \frac{d}{d\overline{t}}(\Delta_{p_1p_1} - \Delta_{p_2p_2}) &= -2\overline{\gamma}_p
(\Delta_{p_1p_1} - \Delta_{p_2p_2}) - \sqrt{2}\overline{\lambda}_+ (\Delta_{p_1R_1} +
\Delta_{p_2R_1}) - \sqrt{2}\overline{\lambda}_- (\Delta_{p_1R_2} - \Delta_{p_2R_2}) \nonumber \\
    &\quad + 2\overline{\gamma}_p \Delta \mathcal{T} \, , \\
    \frac{d}{d\overline{t}}\Delta_{R_1R_2} &= \frac{1}{\sqrt{2}} (\Delta_{p_1R_1} + \Delta_{p_2R_1})
+ \frac{1}{\sqrt{2}} (\Delta_{p_1R_2} - \Delta_{p_2R_2}) - (\overline{\lambda}_+ +
\overline{\lambda}_-) \overline{\gamma}_q \Delta_{R_1R_2} \nonumber \\
    &\quad + \overline{\gamma}_q \Delta \mathcal{T} \, ,
\end{align}
which, as can be seen, are coupled to the momentum-position correlations:
\begin{align}
    \frac{d}{d\overline{t}}(\Delta_{p_1R_1} + \Delta_{p_2R_1}) &= -\sqrt{2}\overline{\lambda}_-
\Delta_{R_1R_2} - \overline{\gamma}_p (\Delta_{p_1R_1} + \Delta_{p_2R_1}) +
\frac{1}{\sqrt{2}}(\Delta_{p_1p_1} - \Delta_{p_2p_2}) \nonumber \\
    &\quad - \overline{\lambda}_+ \overline{\gamma}_q (\Delta_{p_1R_1} + \Delta_{p_2R_1}) \, , \\
    \frac{d}{d\overline{t}}(\Delta_{p_1R_2} - \Delta_{p_2R_2}) &= -\sqrt{2}\overline{\lambda}_+
\Delta_{R_1R_2} - \overline{\gamma}_p (\Delta_{p_1R_2} - \Delta_{p_2R_2}) +
\frac{1}{\sqrt{2}}(\Delta_{p_1p_1} - \Delta_{p_2p_2}) \nonumber \\
    &\quad - \overline{\lambda}_- \overline{\gamma}_q (\Delta_{p_1R_2} - \Delta_{p_2R_2}) \, .
\end{align}

To diagonalize the interaction between the momentum-position correlations, we introduce the
variables $u$ and $v$:
\begin{align}
    u &:= \frac{1}{\sqrt{2}} \left[ (1+2\chi)(\Delta_{p_1R_1} + \Delta_{p_2R_1}) + (\Delta_{p_1R_2}
- \Delta_{p_2R_2}) \right] \, , \\
    v &:= \frac{1}{\sqrt{2}} \left[ -(1+2\chi)(\Delta_{p_1R_1} + \Delta_{p_2R_1}) + (\Delta_{p_1R_2}
- \Delta_{p_2R_2}) \right] \, .
\end{align}
Using these variables, the reduced differential equations can be rewritten in a
compact matrix
form:
\begin{equation}
\frac{d\boldsymbol{Y}}{d\overline{t}} = \mathbf{M}_{\mathrm{sub}} \boldsymbol{Y} + \mathbf{S}
\, ,
\label{eq:reduced_matrix_eq}
\end{equation}
where $\boldsymbol{Y} = (\Delta_{p_1p_1} - \Delta_{p_2p_2}, \Delta_{R_1R_2}, u, v)^T$,
\begin{equation}
    \mathbf{M}_{\mathrm{sub}} = \begin{pmatrix}
        -2\overline{\gamma}_p & 0 & -2 & 0 \\
        0 & -2(1+\chi)\overline{\gamma}_q & \frac{1+\chi}{1+2\chi} & \frac{\chi}{1+2\chi} \\
        (1+\chi) & -2(1+2\chi) & -\overline{\gamma}_p (1+(1+2\chi)(1+\delta)) & \overline{\gamma}_p
\chi (1+\delta) \\
        -\chi & 0 & \overline{\gamma}_p \chi (1+\delta) & -\overline{\gamma}_p
(1+(1+\chi)(1+\delta))
    \end{pmatrix} \, ,
\end{equation}
and the source row vector is given by
\begin{equation}
    \mathbf{S} = \Delta\mathcal{T} (
        2\overline{\gamma}_p,  
        \overline{\gamma}_q, 
        0,
        0
    )^T  \, ,
\end{equation}
This reduced system allows us to solve for the steady state analytically.

\subsection{Analytical Derivation of Linear Thermal Response}

We explicitly solve the linear system for the steady state:
\begin{equation}
\frac{d\boldsymbol{Y}}{d\overline{t}} = \mathbf{M}_{\mathrm{sub}} \boldsymbol{Y} + \mathbf{S} = 0 \,
,
\end{equation}
leading to the following four equations:
\begin{align}
    & -2\overline{\gamma}_p (\Delta_{p_1p_1} - \Delta_{p_2p_2}) - 2 u = -2\overline{\gamma}_p \Delta
\mathcal{T} \, , \label{eq:ss1} \\
    & -2(1+\chi)\overline{\gamma}_q \Delta_{R_1R_2} + \frac{1+\chi}{1+2\chi} u +
\frac{\chi}{1+2\chi} v = -\overline{\gamma}_q \Delta \mathcal{T} \, , \label{eq:ss2} \\
    & (1+\chi)(\Delta_{p_1p_1} - \Delta_{p_2p_2}) - 2(1+2\chi) \Delta_{R_1R_2} - \overline{\gamma}_p
(1 + (1+2\chi)(1+\delta)) u + \overline{\gamma}_p \chi (1+\delta) v = 0 \, , \label{eq:ss3} \\
    & -\chi (\Delta_{p_1p_1} - \Delta_{p_2p_2}) + \overline{\gamma}_p \chi (1+\delta) u -
\overline{\gamma}_p ( 1 + (1+\chi)(1+\delta)) v = 0\, . \label{eq:ss4}
\end{align}


From Eq.~\eqref{eq:ss1}, we immediately obtain a relation between $u$ and $\Delta_{p_1p_1} -
\Delta_{p_2p_2}$:
\begin{equation}
    u = \overline{\gamma}_p \{\Delta \mathcal{T} - (\Delta_{p_1p_1} - \Delta_{p_2p_2}) \}\, .
\label{eq:u_sol}
\end{equation}
Next, we substitute Eq.~\eqref{eq:u_sol} into Eq.~\eqref{eq:ss4} to find $v$ in terms of
$\Delta_{p_1p_1} - \Delta_{p_2p_2}$:
\begin{equation}
    v = \frac{\chi}{1 + (1+\chi)(1+\delta)} \left[ 
    \overline{\gamma}_p (1+\delta) \Delta \mathcal{T} - 
    \left\{ \frac{1}{\overline{\gamma}_p} + \overline{\gamma}_p (1+\delta) \right\} (\Delta_{p_1p_1}
- \Delta_{p_2p_2}) \right] \, .
    \label{eq:v_sol}
\end{equation}
Substituting $u$ and $v$ into Eq.~\eqref{eq:ss2}, we can express $\Delta_{R_1R_2}$ in terms of
$\Delta_{p_1p_1} - \Delta_{p_2p_2}$.
Finally, substituting all variables into Eq.~\eqref{eq:ss3}
yields the linear relation: 
\begin{equation}
    \Delta_{p_1p_1} - \Delta_{p_2p_2} = \mathcal{C}_{y1} \Delta \mathcal{T}
    \label{eq:y1_analytical} \, ,
\end{equation}
where
\begin{equation}
    \mathcal{C}_{y1} = \left[ 1 + \frac{\chi^2}{\overline{\gamma}_p^2 (1+\chi) \{1 +
2(1+\chi)(1+\delta) + (1+2\chi)(1+\delta)^2\}} \right]^{-1} \, .
    \label{eq:Cy1_def}
\end{equation}
The constant $\mathcal{C}_{y1}$ is strictly positive, according to
Eqs.~(\ref{eqn:positive_lambda}) and (\ref{eqn:gamma_delta}), and satisfies the inequality $0 <
\mathcal{C}_{y1} < 1$.
This implies that for a positive temperature difference $\Delta \mathcal{T} > 0$,
we have $0 < \Delta_{p_1p_1} - \Delta_{p_2p_2} < \Delta \mathcal{T}$.
We thus find
\begin{equation}
    \Delta \mathcal{T} - (\Delta_{p_1p_1} - \Delta_{p_2p_2}) = (1 - \mathcal{C}_{y1}) \Delta
\mathcal{T} > 0 \, .
    \label{eq:diff_pos}
\end{equation}
Similarly, we can show that the steady state solution for $\Delta_{R_1 R_2}$ is linearly proportional to $\Delta \mathcal{T}$:
\begin{align}
\Delta_{R_1 R_2} 
&=  \frac{1}{2(1+2\chi)} 
\frac{1}{(1+\chi) \{
1 + 2 (1+\chi)(1+\delta) +(1+2\chi) (1+\delta)^2\} + \chi^2/\overline{\gamma}^2_p}
\nonumber \\
& 
\times \left[
- 
\chi^2
 + \left\{
\chi (1+\delta) - \frac{1+\chi}{\chi} (1+(1+\chi)(1+\delta))
\right\} \frac{1}{1 + (1+\chi) (1+\delta)} \right. \nonumber \\
&\left. \times \left\{
- \chi (1+\chi) \{
1 + 2 (1+\chi)(1+\delta) +(1+2\chi) (1+\delta)^2\}
\right. \right. \nonumber \\
& \left. \left. 
+ \chi \left( 1 + \delta \right)\chi^2
\right\}
\right] \Delta \mathcal{T} \label{eqn:sol_dr1r2_app} \, .
\end{align}

We are now ready to find the Fourier-type linear thermal response. 
Using the first law
\eqref{eq:first_law_avg}, the heat current from the heat bath at $T_1$ is
re-expressed as
\begin{align}
J_1 (t)
&= \frac{1}{2} \left\{ \left\lceil   \frac{d\tilde{Q}_{1,t}}{dt} \right\rfloor  - \left\lceil
\frac{d\tilde{Q}_{2,t}}{dt} \right\rfloor  \right\}
+ \frac{1}{2} \frac{d}{dt} \left\lceil H \right\rfloor
\nonumber \\
&= \frac{1}{2} (J_1-J_2) + \frac{1}{2} \frac{d}{dt} \left\lceil  H \right\rfloor \, .
\label{eqn:j1_app}
\end{align}
In the steady state, $ \frac{d}{dt} \left\lceil  H \right\rfloor =0$.
Therefore,
using the expression in (\ref{eq:diff_currents}) along with the expressions for the
correlations just derived,
the steady heat current from the heat bath at $T_1$ is expressed as a linear function of
$\Delta \mathcal{T}$:
\begin{equation}
    J_1^\text{ss} = \kappa \Delta \mathcal{T} \, ,
\label{eqn:fourier_general}
\end{equation}
for the thermal conductivity $\kappa$ defined by
\begin{align}
    \kappa &= \frac{p_0^2}{m t_0} \frac{\overline{\gamma}_p}{2} \Bigg[ (1 - \mathcal{C}_{y1}) \left[
1 + (1+\delta)\overline{\gamma}_p^2 \left\{ 1 +
\frac{1+\chi+(1+2\chi)(1+\delta)}{1+(1+\chi)(1+\delta)}(1+\delta) \right\} + (1+\delta)(1+\chi)
\right] \nonumber \\
    &\quad + \mathcal{C}_{y1} \frac{\chi^2 (1+\delta)^2}{1+(1+\chi)(1+\delta)} \Bigg] \ge 0\, .
    \label{eq:J1_final_linear}
\end{align}
Since $0 < \mathcal{C}_{y1} < 1$ as shown in Eq.~\eqref{eq:Cy1_def}, both coefficients
$(1-\mathcal{C}_{y1})$ and $\mathcal{C}_{y1}$ are positive. 
Furthermore, all parameters inside the
brackets are positive due to Eq.~(\ref{eqn:para_chi}).
Thus, $\kappa$ is strictly positive, analytically confirming that the heat flows from the hotter to
the colder bath, satisfying Fourier-type linear thermal response.

It should be noted that while we refer to this linear relationship as Fourier's law, in the strict
sense, it describes the local proportionality between heat flux and
temperature gradient in continuum mechanics.
In the context of microscopic models for finite-size systems, confirming the linear dependence of
the heat current on the temperature difference is the standard procedure to verify the validity of
thermal conduction behavior \cite{RLL1967}.

Let us consider the standard Brownian motion limit, which corresponds to $\gamma_q \to 0$
(or equivalently $\delta \to -1$).
The thermal conductivity $\kappa$ given above then simplifies to
\begin{align}
   \kappa(\delta = -1) &= \frac{p_0^2}{m t_0} \frac{\overline{\gamma}_p}{2} \Bigg[ 1 -  \left( 1 + \frac{\chi^2}{\overline{\gamma}_p^2 (1+\chi) } \right)^{-1}  \Bigg]  \ge 0 \, .
\end{align}
In Ref.~\cite{Sekimoto2010}, the heat conduction of an $N=2$ network model was investigated using an overdamped harmonic system where inertial effects are ignored. 
It was shown that the heat current of their model diverges in the limit
$K_\text{int} \to  \infty$, which was considered a fundamental limitation of the model without the inertial effect.
In contrast, this equation shows that $\kappa(\delta = -1)$ remains finite even in the limit $K_\text{int} \to \infty$ (infinite $\chi$). Thus, our model successfully resolves the divergence problem present in Ref.~\cite{Sekimoto2010}.
Now we consider the infinite limit of $\overline{\gamma}_p$. 
The heat conductivity $\kappa(\delta = -1)$ vanishes and this behavior is consistent with the result in Ref.~\cite{Sekimoto2010}. 
For $\delta \neq -1$, the heat conductivity in the same limit is
\begin{equation}
  \kappa \simeq \frac{p_0^2}{m t_0} \frac{\overline{\gamma}_p}{2} \frac{\chi^2 (1+\delta)}{1+\chi} \, ,
\end{equation}
and hence diverges.
This is a qualitative difference between standard Brownian motion and our generalized model.

\subsection{Physical Interpretation: Energy Equipartition and Thermal Contact Resistance}

To understand the physical consistency of our model, let us consider the non-interacting limit,
$K_\text{int} \to 0$, which corresponds to $\chi \to 0$. In this limit, the two oscillators should
decouple and equilibrate with their respective heat baths independently, satisfying the energy
equipartition law.
Substituting $\chi = 0$ into Eq.~\eqref{eq:y1_analytical}, we find that the coefficient
$\mathcal{C}_{y1}$ becomes 1, leading to
\begin{equation}
   \Delta_{p_1p_1} - \Delta_{p_2p_2} = \Delta \mathcal{T} \, .
\end{equation}
Recalling the definitions of the adimensional variables, $\Delta_{p_1 p_1} - \Delta_{p_2 p_2}$ and
$\Delta \mathcal{T}$, this equation gives
\begin{equation}
    \left\lceil  \tilde{p}_{1,t}^2 \right\rfloor - \left\lceil \tilde{p}_{2,t}^2 \right\rfloor= m
k_B (T_1 - T_2) \, .
\end{equation}
This implies the law of equipartition of energy, $\left\lceil \tilde{p}_{i,t}^2\right\rfloor/2m =
k_B T_i / 2$.
Thus, our model correctly reproduces equilibrium statistical mechanics in the non-interacting limit.

In the presence of interaction ($\chi > 0$), we found in Eq.~\eqref{eq:y1_analytical} that
$\Delta_{p_1p_1} - \Delta_{p_2p_2} < \Delta \mathcal{T}$, leading to
\begin{equation}
    \frac{\left\lceil\tilde{p}_{1,t}^2\right\rfloor  - \left\lceil\tilde{p}_{2,t}^2\right\rfloor }{m
k_B} < T_1 - T_2 \, . \label{eqn:p^2-=T-}
\end{equation}
Let us define the effective temperature $T_{\text{kin}, i}$ of the $i$-th harmonic oscillator by
\begin{equation}
T_{\text{kin}, i} := \frac{\left\lceil\tilde{p}_{i,t}^2\right\rfloor}{m k_B} \, .
\end{equation}
Here, we used the fact that the expectation values of the kinetic and potential energies of harmonic
oscillators are the same.
Using this, Eq.\ (\ref{eqn:p^2-=T-}) is re-expressed as
\begin{equation}
    T_{\text{kin}, 1} - T_{\text{kin}, 2} < T_1 - T_2 \, .
\end{equation}
This reduction in the effective temperature difference indicates that the Brownian particles do not 
thermalize with their respective heat bath temperatures due to the finite coupling at the
boundaries.
This leads to a temperature discontinuity, or a ``temperature jump'', at the interface between the
heat bath and the system.
This phenomenon is a microscopic manifestation of thermal boundary resistance, analogous to Kapitza
resistance in the context of interfacial heat transport \cite{Lepri2003,Swartz1989,Giri2020}.
Our generalized Brownian motion model naturally captures this non-equilibrium feature characteristic
of finite-size systems.

\section{Time Evolution and Transient Dynamics}
\label{sec:transient}

To gain further insight into the non-equilibrium behavior of the system, we investigate the time
evolution of the heat currents.
To this end, we numerically solve the coupled differential equations derived in
Sec.~\ref{sec:steady_state}.
Initially, we consider two non-interacting harmonic oscillators in thermal equilibrium
with their respective heat baths at temperatures $T_1$ and $T_2$, that is,
we set $K_\text{int} = 0$ in Eq.(\ref{eq:interaction}), and 
the initial phase space distribution, a product of individual phase space distributions
for each oscillator is given by
\begin{equation}
    \rho_{\blue{\text{ini}}}(q_1, p_1, q_2, p_2) = \prod_{i=1}^2 \frac{1}{Z_i} \exp\left[ -\beta_i \left(
\frac{p_i^2}{2m_i} + \frac{K_i (q_i - x_{(i)})^2}{2} \right) \right] \, ,
    \label{eq:initial_distribution}
\end{equation}
where $Z_i=2\pi/\overline{\beta}_i$ is the partition function for the $i$-th oscillator.
For $t\ge 0$, the initial state will evolve according to the Hamiltonian
Eq.~(\ref{eq:hamiltonian}) with the interaction term in Eq.~(\ref{eq:interaction}) for
$K_\text{int} \ne 0$.

Based on this distribution, the initial conditions for the correlation functions are determined. 
For the simplified case ($m_i=m, K_i=K, x_{(i)}=0$), they correspond to
\begin{align}
    \Delta_{p_i p_i} &= \frac{1}{\overline{\beta}_i} \, , \\ 
    \Delta_{R_1 R_1} = \Delta_{R_2 R_2} &= \frac{1}{2\overline{\beta}_1} +
\frac{1}{2\overline{\beta}_2} \, ,\\
    \Delta_{R_1 R_2} &= \frac{1}{2\overline{\beta}_1} - \frac{1}{2\overline{\beta}_2} \, ,
    \label{eq:initial_correlations}
\end{align}
and all other off-diagonal correlations are zero.

Before showing numerical results, it is instructive to check the initial heat current analytically. 
By substituting the initial correlations (representing the state at $t=0$) into the general
expression for the heat current in(Eq.~\eqref{eq:mean_heat_current},
we can explicitly derive the
instantaneous heat current $J_1(0)$ at the moment the interaction is turned on.
For the generalized model ($\gamma_q \neq 0$), the initial heat current is given by
\begin{equation}
    J_1(0) = -\frac{p_0^2}{m t_0} \overline{\gamma}_p (1+\delta) \chi \left(
\frac{1+\chi}{\overline{\beta}_1} + \frac{\chi}{\overline{\beta}_2} \right) \, .
    \label{eq:J1_initial_analytical}
\end{equation}
This expression clearly shows that if $\gamma_q \neq 0$ (and thus $\delta \neq -1$), $J_1(0)$
generally does not vanish.
The sign of the initial current depends on the interaction parameter $\chi$ constrained by Eq.~(\ref{eqn:para_chi}).
For attractive interaction ($\chi > 0$ or $K_\text{int} > 0$), the current
$J_1(0)$ is negative, indicating heat outflow.
From the symmetry of the system, $J_2(0)$ is obtained by exchanging the indices 1 and 2:
\begin{equation}
    J_2(0) = -\frac{p_0^2}{m t_0} \overline{\gamma}_p (1+\delta) \chi \left(
\frac{\chi}{\overline{\beta}_1} + \frac{1+\chi}{\overline{\beta}_2} \right) \, .
    \label{eq:J2_initial_analytical}
\end{equation}
Then, the initial rate of change of the system energy is determined by the sum of these currents:
\begin{equation}
    \left. \frac{d\left\lceil H \right\rfloor}{dt} \right|_{t=0} = J_1(0) + J_2(0) = -\frac{p_0^2}{m
t_0} \overline{\gamma}_p (1+\delta) \chi (1+2\chi) \left( \frac{1}{\overline{\beta}_1} +
\frac{1}{\overline{\beta}_2} \right) \, .
\end{equation}
This quantity reflects the system's immediate relaxation response to the sudden change in potential
energy caused by introducing the interaction.

\subsection{Differentiable and Non-differentiable Trajectories}

\begin{figure}[ht!]
\includegraphics[scale=0.2]{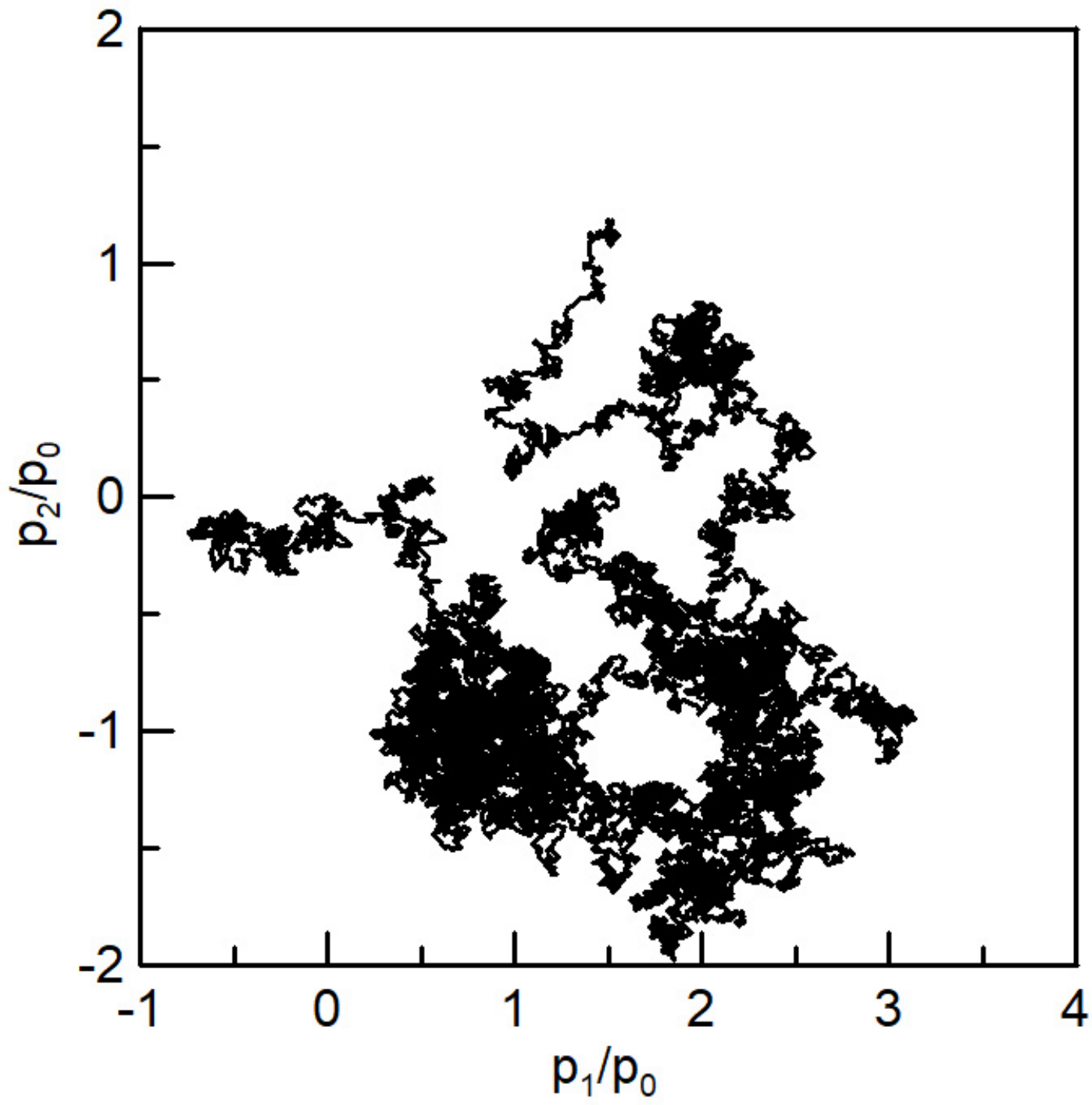}
\hspace{2cm}
\includegraphics[scale=0.2]{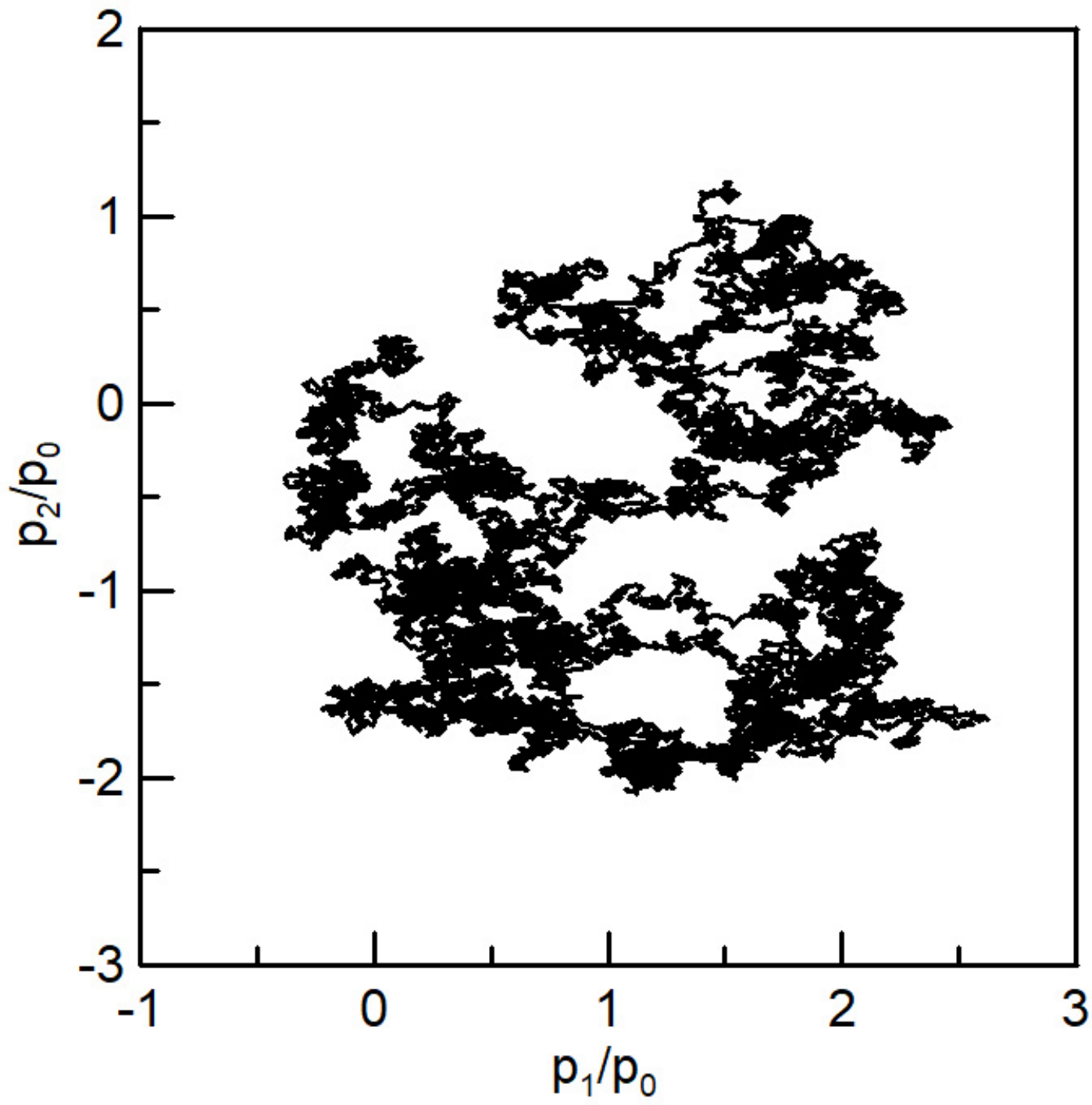}\\
\includegraphics[scale=0.2]{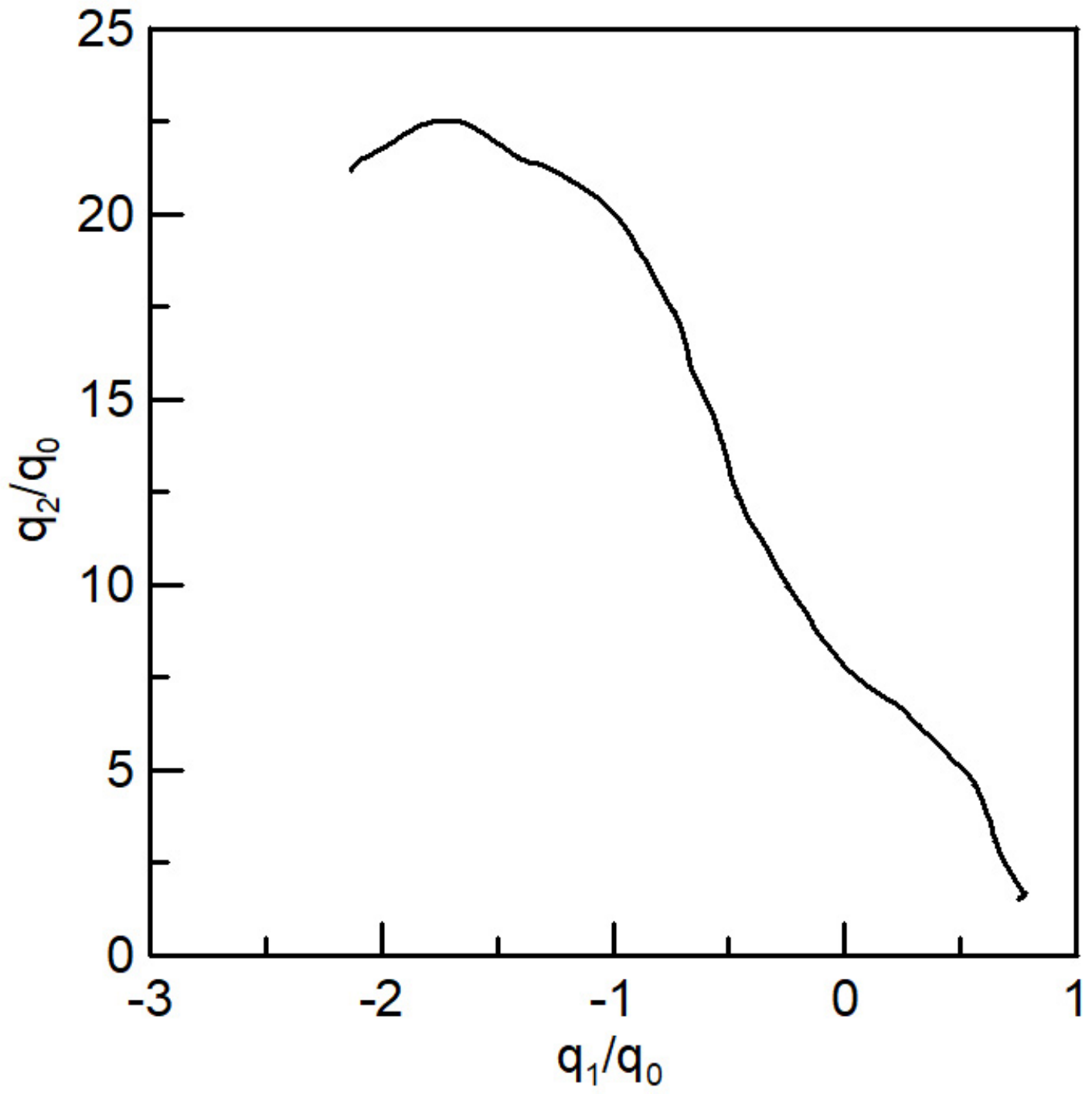}
\hspace{2cm}
\includegraphics[scale=0.2]{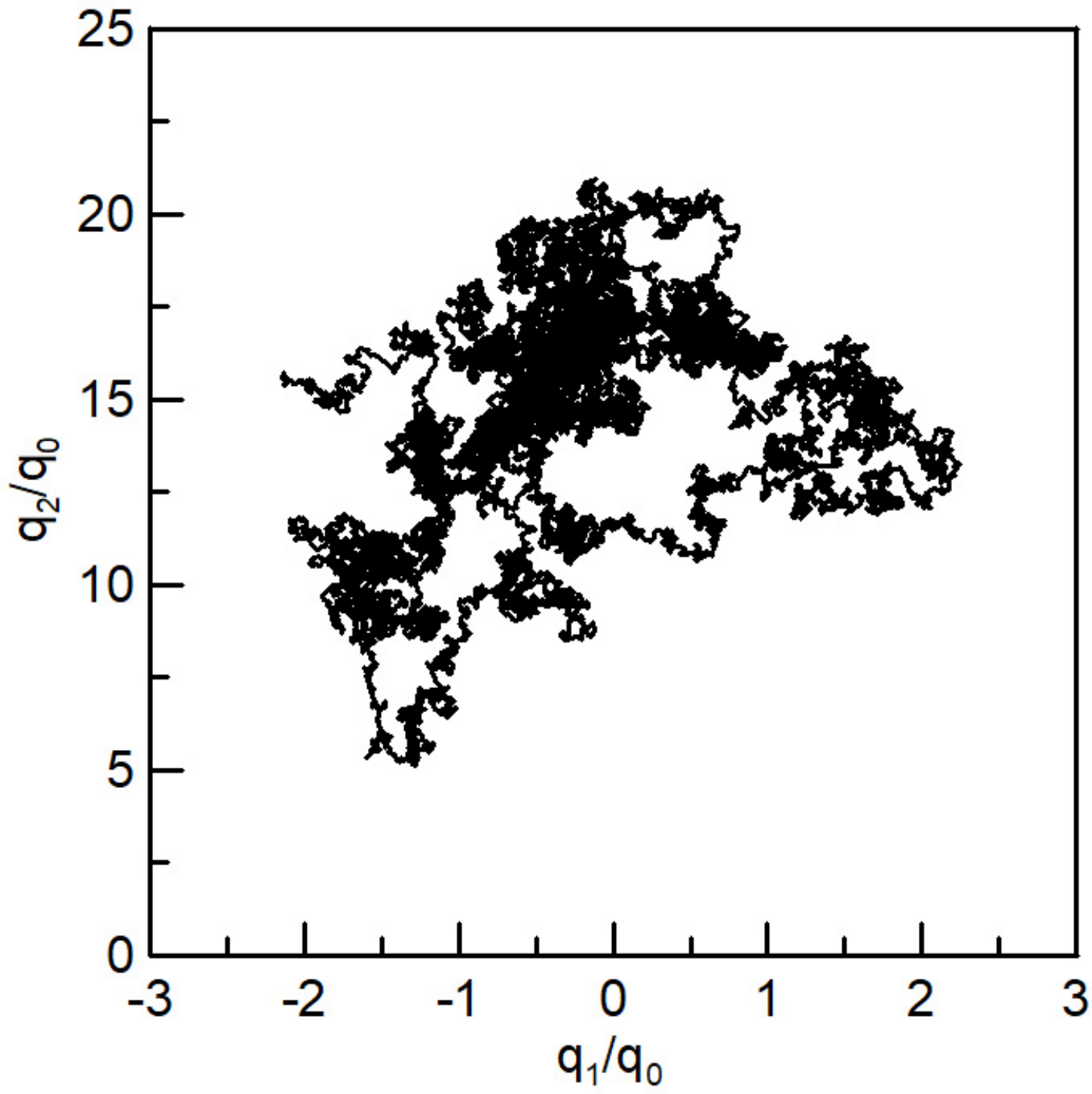}
\caption{Trajectories obtained by solving Eqs.\ (\ref{eq:gen_dq}) and (\ref{eq:gen_dp}) are
plotted.
The top-left and top-right panels show the momentum trajectories for $\gamma_q = 0$ and $\gamma_q =
1$ in the $(p_1, p_2)$ plane, respectively. Both appear as zigzag lines. The bottom-left and
bottom-right panels show the position trajectories in the $(q_1, q_2)$ plane. For $\gamma_q = 0$
(left panel), the trajectory is smooth, whereas for $\gamma_q = 1$ (right panel), it becomes zigzag
due to the direct influence of the noise term.
}                  
\label{fig:tras}
\end{figure}

To illustrate the role of fluctuation and dissipation in the equation of position, we first examine
the qualitative differences in the trajectories of standard Brownian motion and our generalized
model.
In the case of standard Brownian motion ($\gamma_q = 0$), there is no noise term in the evolution
equation for the position (\ref{eq:gen_dq}).
Consequently, while the momentum trajectory is stochastic and exhibits a zigzag pattern, the
position trajectory remains smooth.
In contrast, for generalized Brownian motion ($\gamma_q \neq 0$), both the momentum and position
trajectories exhibit this zigzag behavior.
This reflects the fundamental property of the Wiener process, where the trajectories are continuous
but nowhere differentiable.
These behaviors can be confirmed numerically in Fig.~\ref{fig:tras}, which displays the trajectories
obtained by solving Eqs.~(\ref{eq:gen_dq}) and (\ref{eq:gen_dp}) with $1/\overline{\beta}_1=2$ and
$1/\overline{\beta}_2=1$.
The top-left and top-right panels show the momentum trajectories in the $(p_1, p_2)$ plane for
$\gamma_q = 0$ and $\gamma_q = 1$, respectively.
Both appear as zigzag lines. 
The bottom-left and bottom-right panels show the corresponding position trajectories in the $(q_1,
q_2)$ plane.
For $\gamma_q = 0$ (left panel), the trajectory is smooth, whereas for $\gamma_q = 1$ (right panel),
it becomes a zigzag line due to the direct influence of the noise term.

\subsection{numerical simulations of thermal conduction}

In the following simulations, we show the time evolutions of heat currents and the change of the
system energy, which are plotted with adimensional quantities.
We introduce adimensional heat currents $j_1$ and $j_2$ by
\begin{align}
j_1 &= \frac{m t_0}{p^2_0} J_1 \, ,\\
j_2 &= \frac{d\epsilon}{d\tau}  - j_1 \, ,
\end{align}
and adimensional form of the change of the system energy is given by
\begin{align}
\frac{d\epsilon}{d\tau} 
=
\frac{mt_0}{p^2_0}
\frac{d}{dt} 
\left\lceil H \right\rfloor \, .
\end{align}

\subsubsection{Standard Brownian Motion ($\overline{\gamma}_q = 0$)}
\label{sec:numerial_sbm}

\begin{figure}[ht!]
\includegraphics[scale=0.2]{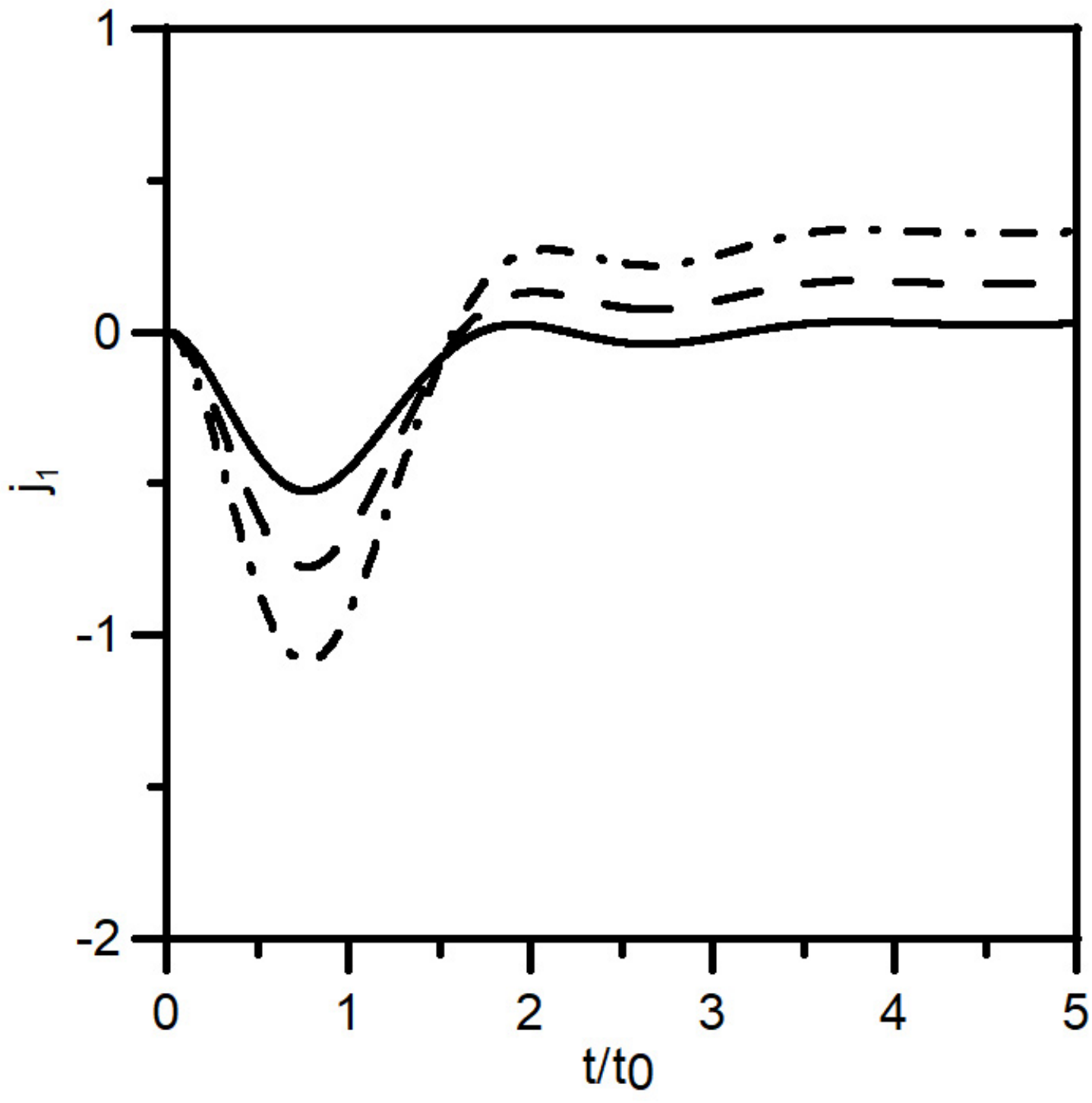}
\includegraphics[scale=0.2]{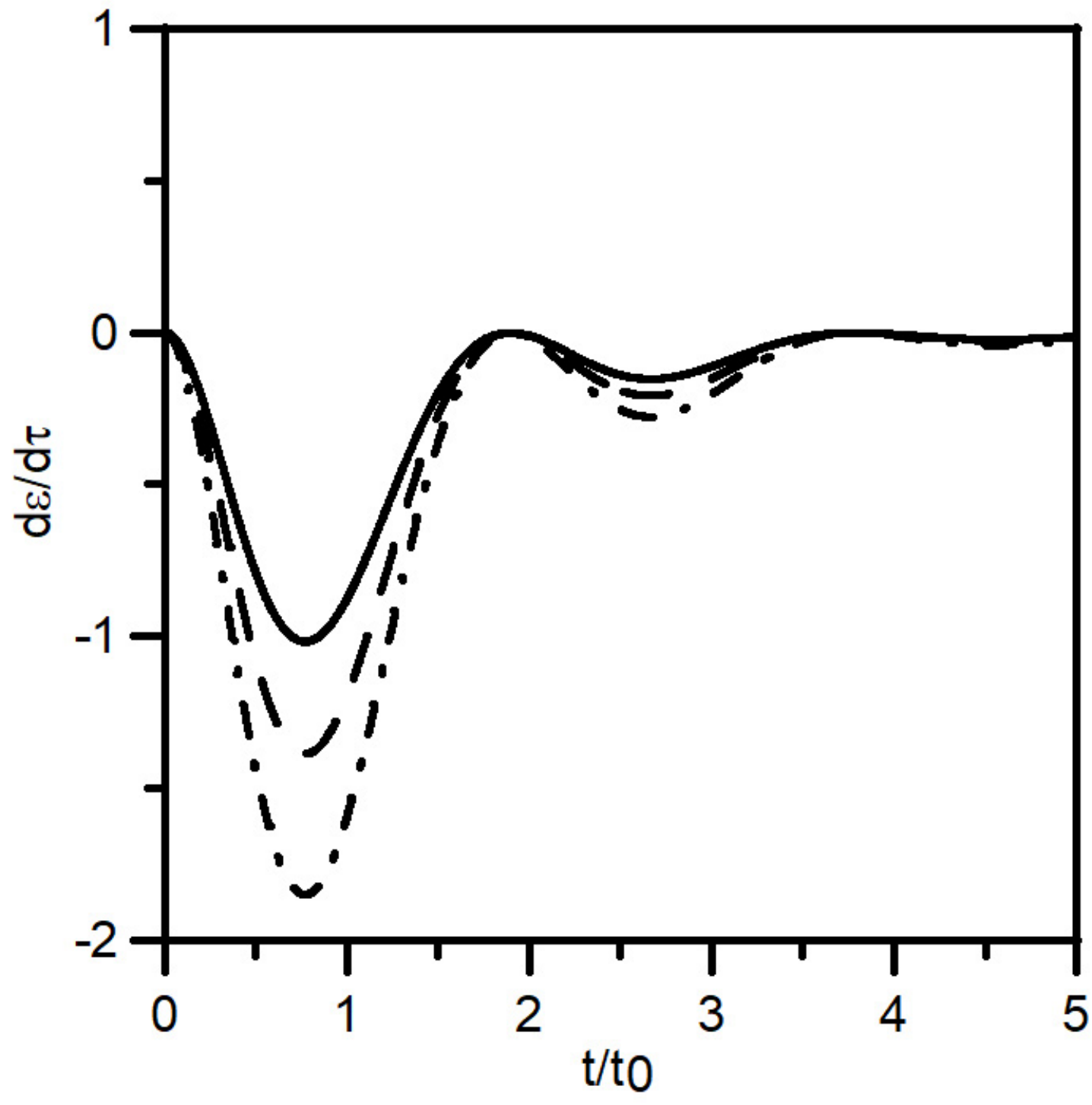}
\includegraphics[scale=0.2]{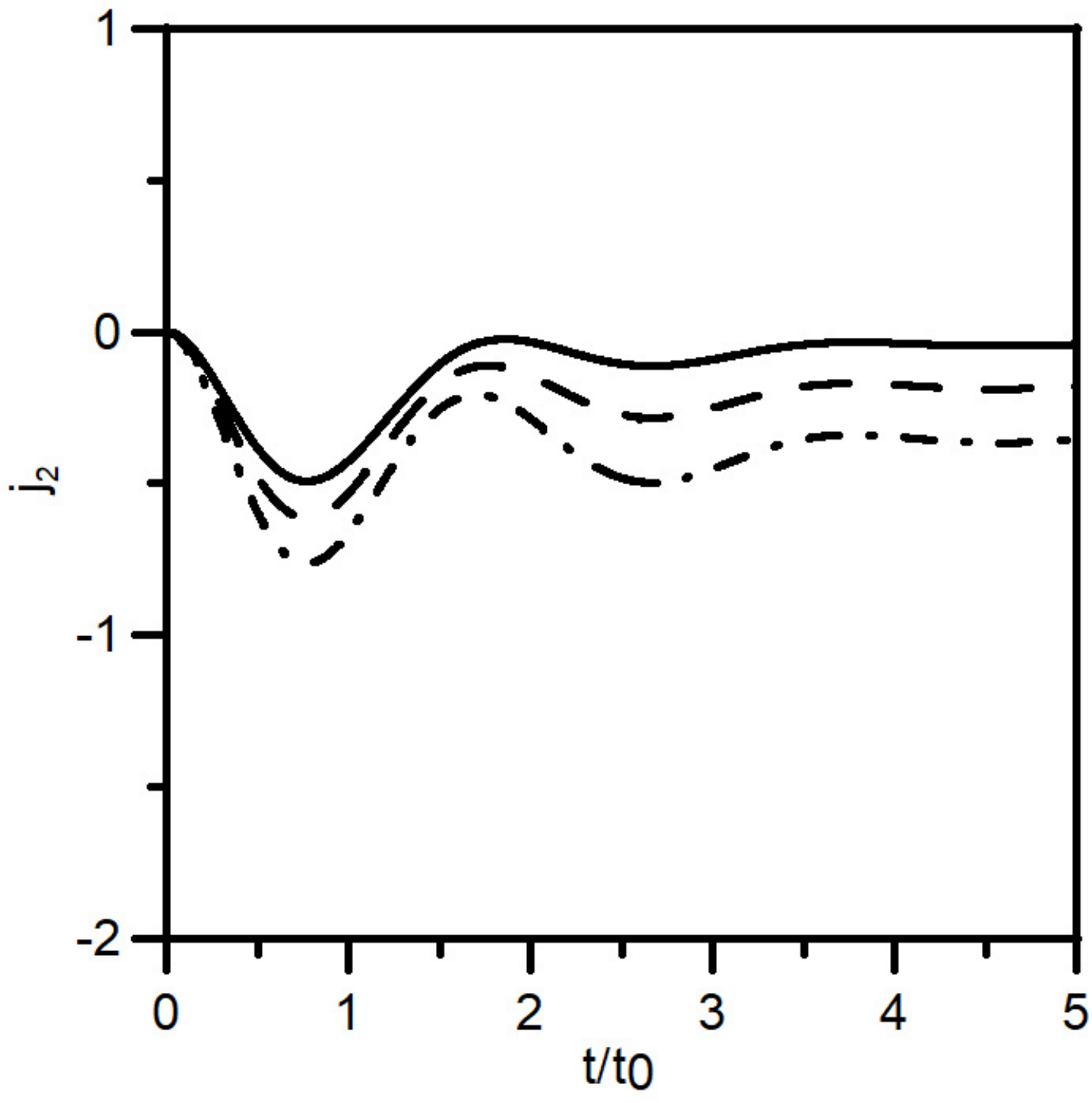}
\caption{
The time evolutions of the heat current $j_1$ (left panel), the change of energy $d\epsilon/d\tau$
(center) and the heat current $j_2$ (right panel) are plotted for the standard Brownian motion,
$\overline{\gamma}_q = 0\,\,\, (\delta =-1)$.
We set $\overline{\gamma}_p = \chi = 1$, $\delta=-1$ and $\overline{\beta}_2=1$. 
The solid, dashed and dash-dotted lines correspond to temperature $1/\beta_1 =1.2$, $2$ and $3$,
respectively.
The three trajectories of $j_1$ converge toward the analytical values obtained through
Eq.~(\ref{eq:J1_final_linear}).
On the center figure, the numerical solutions converge toward zero asymptotically in time, showing
the system reaches a steady state.
}                  
\label{fig:delta=-1}
\end{figure}

First, we discuss the results for standard Brownian motion, which corresponds to the case where
$\overline{\gamma}_q = 0$ (or $\delta = -1$).
In Fig.~\ref{fig:delta=-1}, the time evolution of the heat current $j_1$ (left panel), the rate of
energy change $d\epsilon/d\tau$ (center), and the heat current $j_2$ (right panel) are plotted for
different temperatures.
The solid, dashed, and dash-dotted lines represent the temperatures $1/\overline{\beta}_1 = 1.2$, 
$2$ and $3$, respectively.
We set $\overline{\gamma}_p = \chi = 1$, $\delta=-1$ and $\overline{\beta}_2=1$. 
As shown, the value of $d\epsilon/d\tau$ vanishes in the long-time limit, which is consistent with the definition of a steady state.
We confirmed that all trajectories for $j_1$ converge toward the analytical values derived from
Eqs.~(\ref{eqn:fourier_general}) and (\ref{eq:J1_final_linear}).
The corresponding steady-state values for $j_2$ are obtained through the first law in the steady
state, $j_1 + j_2 = 0$.
In the steady state, heat flows from the hotter bath to the colder one ($J_1 > 0, J_2 < 0$) as
expected, satisfying the energy conservation law $J_1 + J_2 = 0$.

The transient behavior exhibits a peculiar feature. 
Immediately after the interaction is turned on at $t=0$,
both heat currents $J_1$ and $J_2$ become negative.
This means that heat flows \textit{from} the system \textit{into} both heat baths.
This ``backflow'' or dissipation of energy into the baths occurs because the introduction of the
interaction term $V_\text{int}$ at $t=0$ increases the potential energy of the system,
effectively raising the ``effective temperature" of the system above that of the hotter bath.
The system dissipates this excess energy to relax to the new non-equilibrium steady state.
It is also notable that $J_1(0) = J_2(0) = 0$ in the standard model, as the velocity is continuous
and the force from the interaction does not instantly affect the work done by the heat bath on the
momentum.
These can be confirmed even from Eqs.~(\ref{eq:J1_initial_analytical}) and
(\ref{eq:J2_initial_analytical}).

\subsubsection{Generalized Brownian Motion ($\overline{\gamma}_q > 0$)}
\label{sec:numerial_gbm_posi}

\begin{figure}[h!]
\includegraphics[scale=0.2]{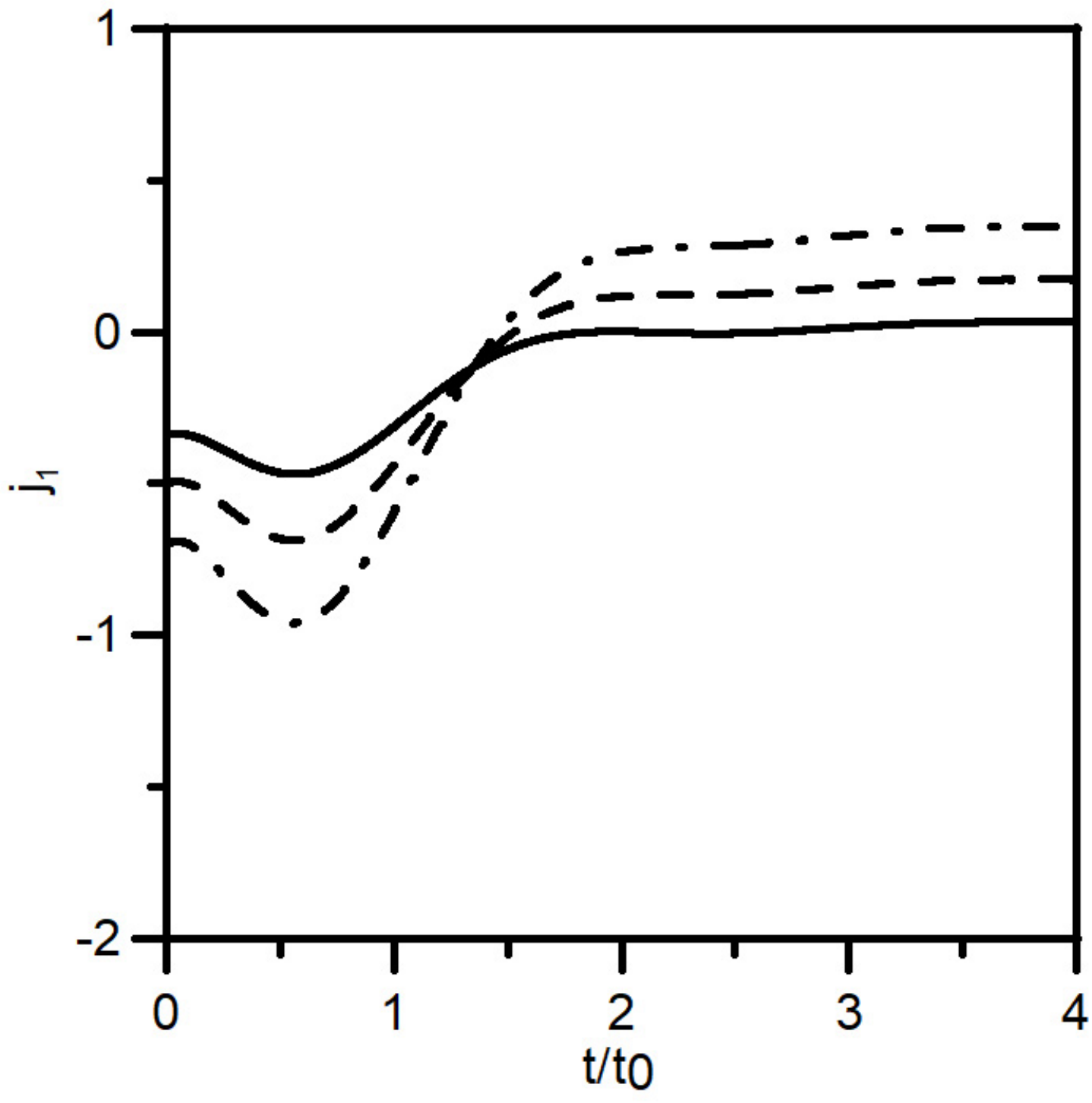}
\includegraphics[scale=0.2]{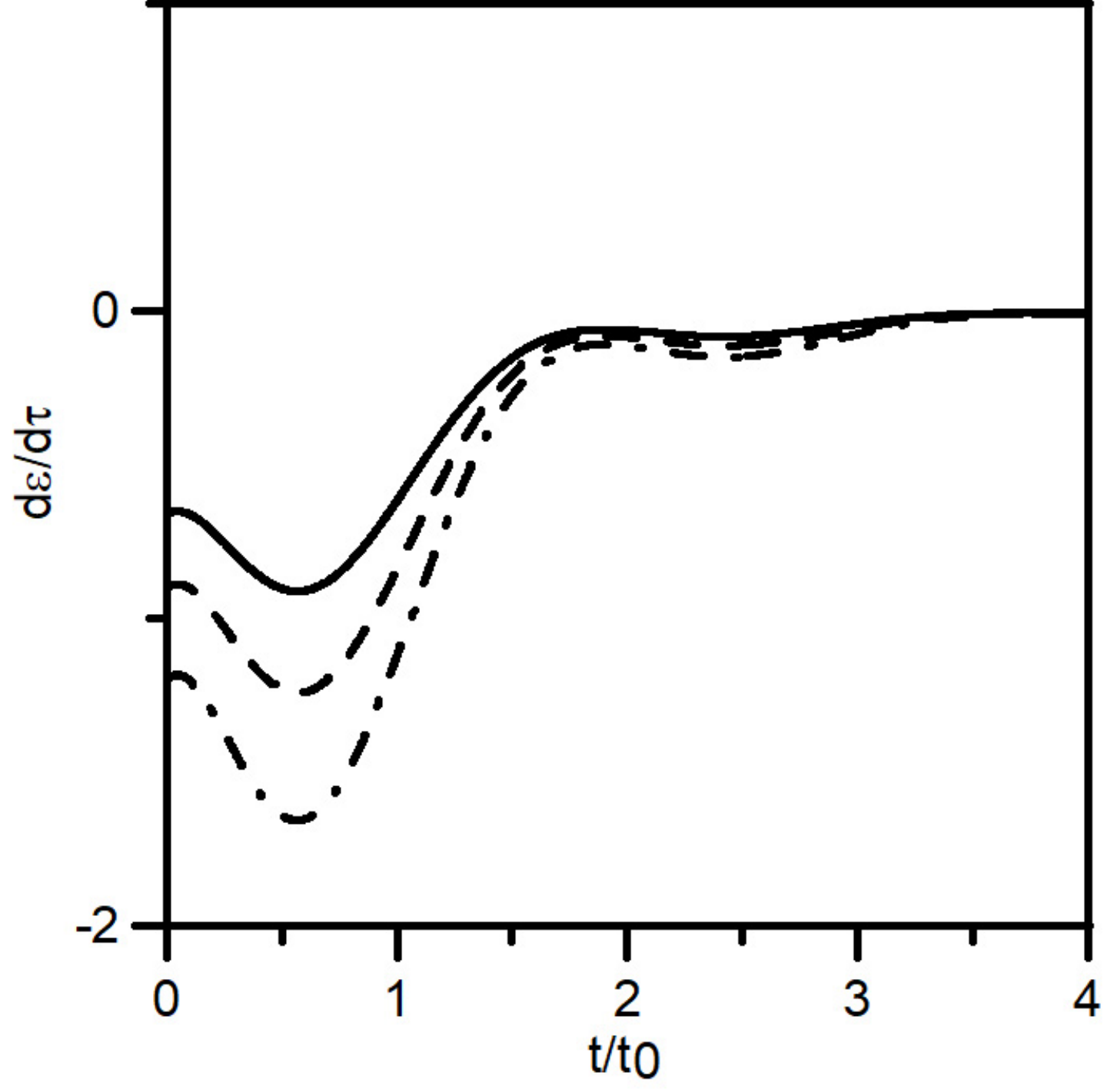}
\includegraphics[scale=0.2]{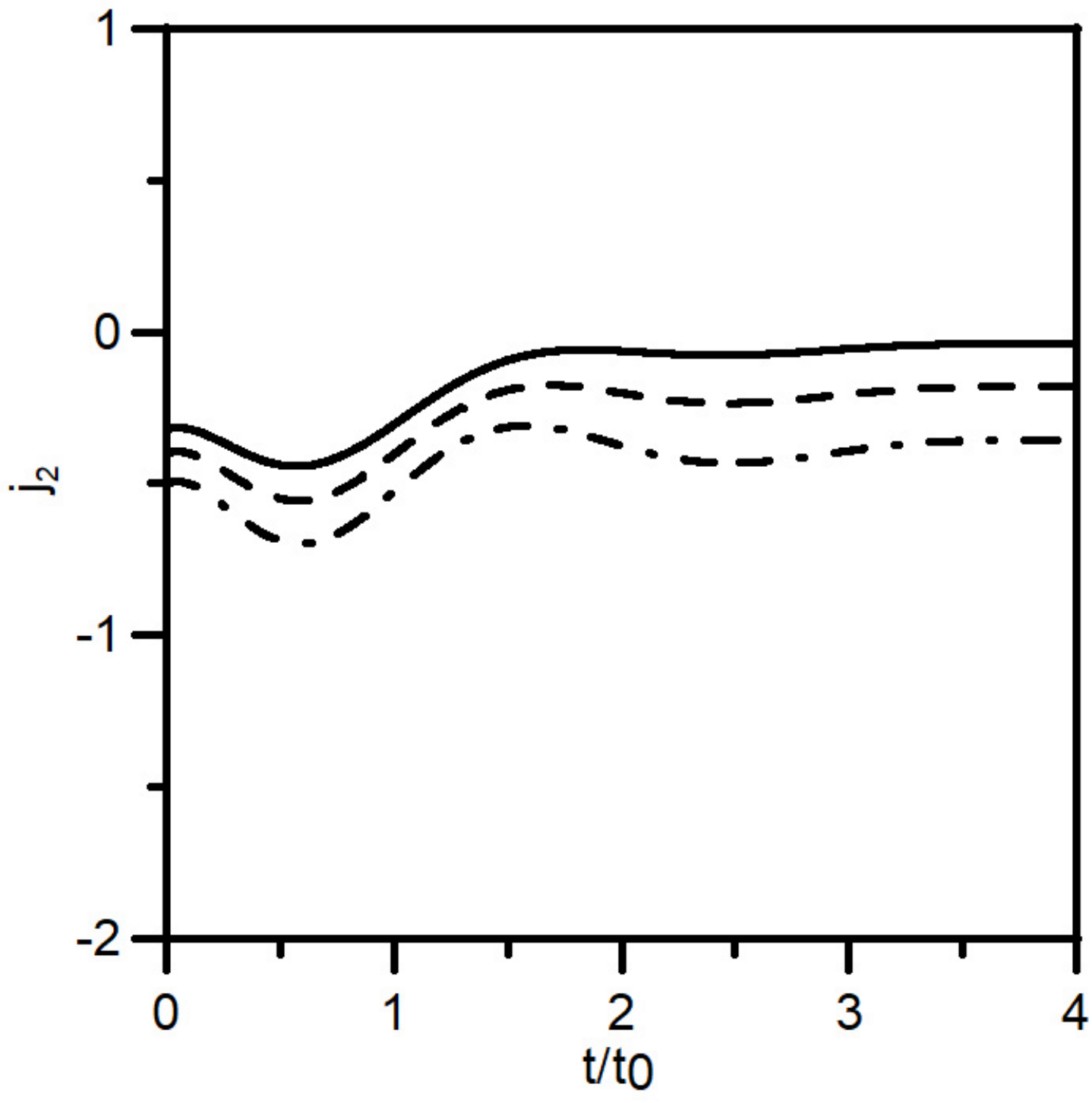}
\caption{
The time evolution of the heat current $j_1$ (left panel), the rate of energy change
$d\epsilon/d\tau$ (center), and the heat current $j_2$ (right panel) for the generalized Brownian
motion with parameter $\overline{\gamma}_q = 0.1$ ($\delta = -0.9$).
We set $\overline{\gamma}_p = \chi = 1$ and $\overline{\beta}_2 = 1$. 
The solid, dashed, and dash-dotted lines correspond to temperatures $1/\overline{\beta}_1 = 1.2$,
$2$, and $3$, respectively.
The three trajectories of $j_1$ converge toward the analytical values obtained through
Eq.~(\ref{eq:J1_final_linear}).
In the center panel, the numerical solutions asymptotically converge toward zero, demonstrating that
the system reaches a steady state.
}                  
\label{fig:delta=-0.9}
\end{figure}

We consider the effect of the position dissipation $\overline{\gamma}_q$ in our generalized model. 
Figure \ref{fig:delta=-0.9} shows the time evolution of the heat current $j_1$ (left panel), the
rate of energy change $d\epsilon/d\tau$ (center), and the heat current $j_2$ (right panel) for the
parameter $\overline{\gamma}_q = 0.1$ ($\delta = -0.9$).
Other parameters are set to be the same as the standard Brownian motion, $\overline{\gamma}_p = \chi
= 1$ and $\overline{\beta}_2 = 1$.
The solid, dashed, and dash-dotted lines represent the temperatures $1/\overline{\beta}_1 = 1.2$,
$2$, and $3$, respectively.
As before, the value of $d\epsilon/d\tau$ vanishes in the long-time limit, which is consistent with
the definition of a steady state.
Furthermore, all trajectories for $j_1$ converge toward the analytical values derived from
Eq.~(\ref{eq:J1_final_linear}).
In the steady state ($t \to \infty$), heat flows from the hotter bath to the colder one ($J_1 > 0,
J_2 < 0$).

A striking difference from standard Brownian motion discussed in Sec.~\ref{sec:numerial_sbm} is
observed at $t=0$.
For $\overline{\gamma}_q > 0$, the heat current starts from a non-zero value, $J_1(0) \neq 0$. 
This instantaneous response is due to the presence of noise and dissipation terms in the position
equation, Eq.~\eqref{eq:gen_dq}.
The sudden change in the interaction potential $V_\text{int}$ at $t=0$ directly affects the heat
current defined in Eq.~\eqref{eq:heat_def} through the term proportional to $\overline{\gamma}_q$.
As predicted analytically in Eq.~\eqref{eq:J1_initial_analytical}, the generalized model allows for
an immediate energy (heat) exchange associated with structural changes in the potential.
It should be noted, however, that this instantaneous jump in the heat current is a consequence of
the idealized assumption that the interaction is turned on as a step function. In any realistic
experimental setup, the coupling would change over a finite time interval.
If the process were modeled using a continuous time-dependent coupling $K_\text{int}(t)$ starting
from $0$, the parameter $\chi$ would also start from $0$ in Eqs.~\eqref{eq:J1_initial_analytical}
and \eqref{eq:J2_initial_analytical}, and these jumps would disappear.
Therefore, this response is an artifact of our idealized setup and would not be observed in more
realistic continuous scenarios.

\begin{figure}[ht!]
\includegraphics[scale=0.2]{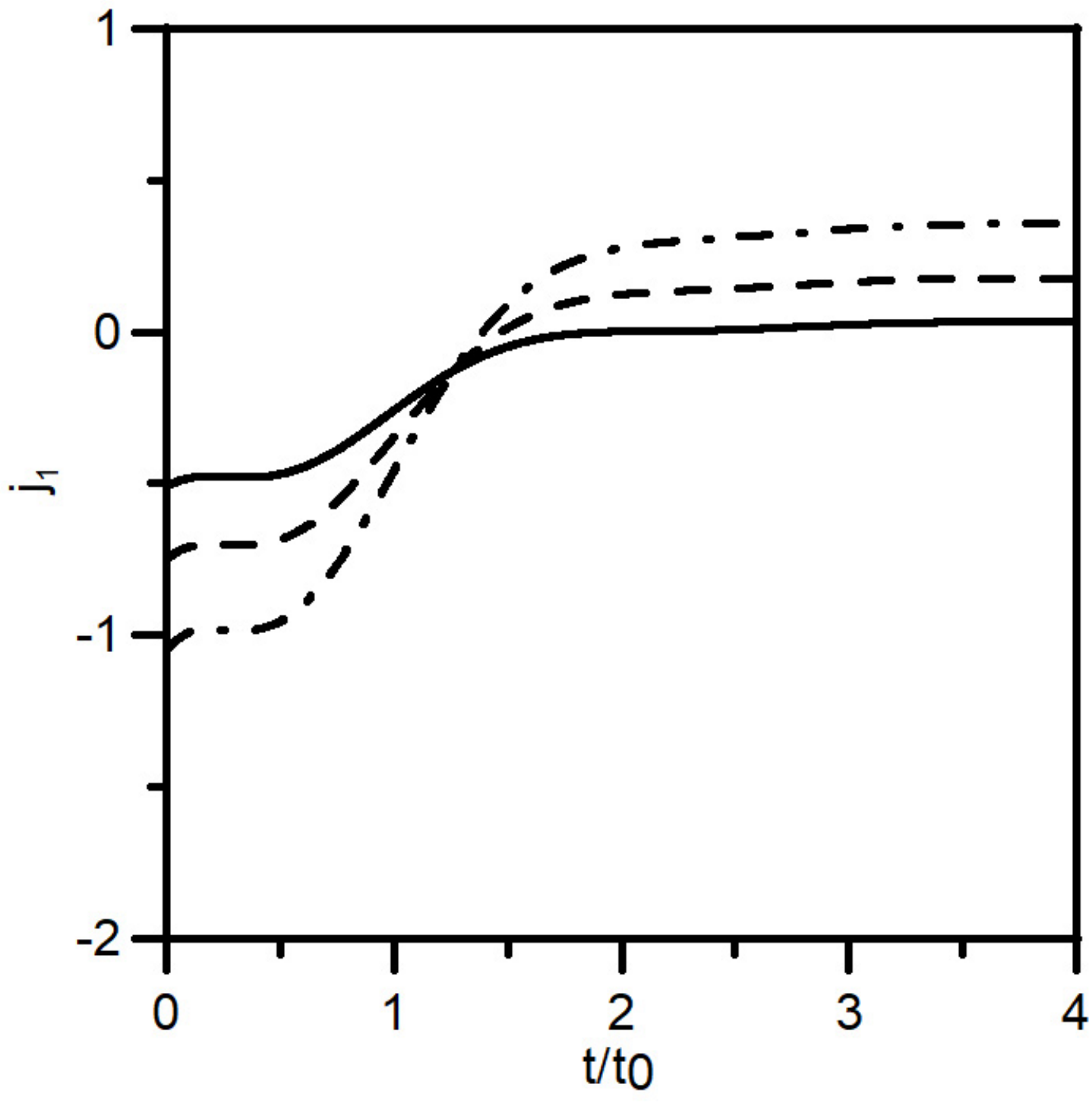}
\includegraphics[scale=0.2]{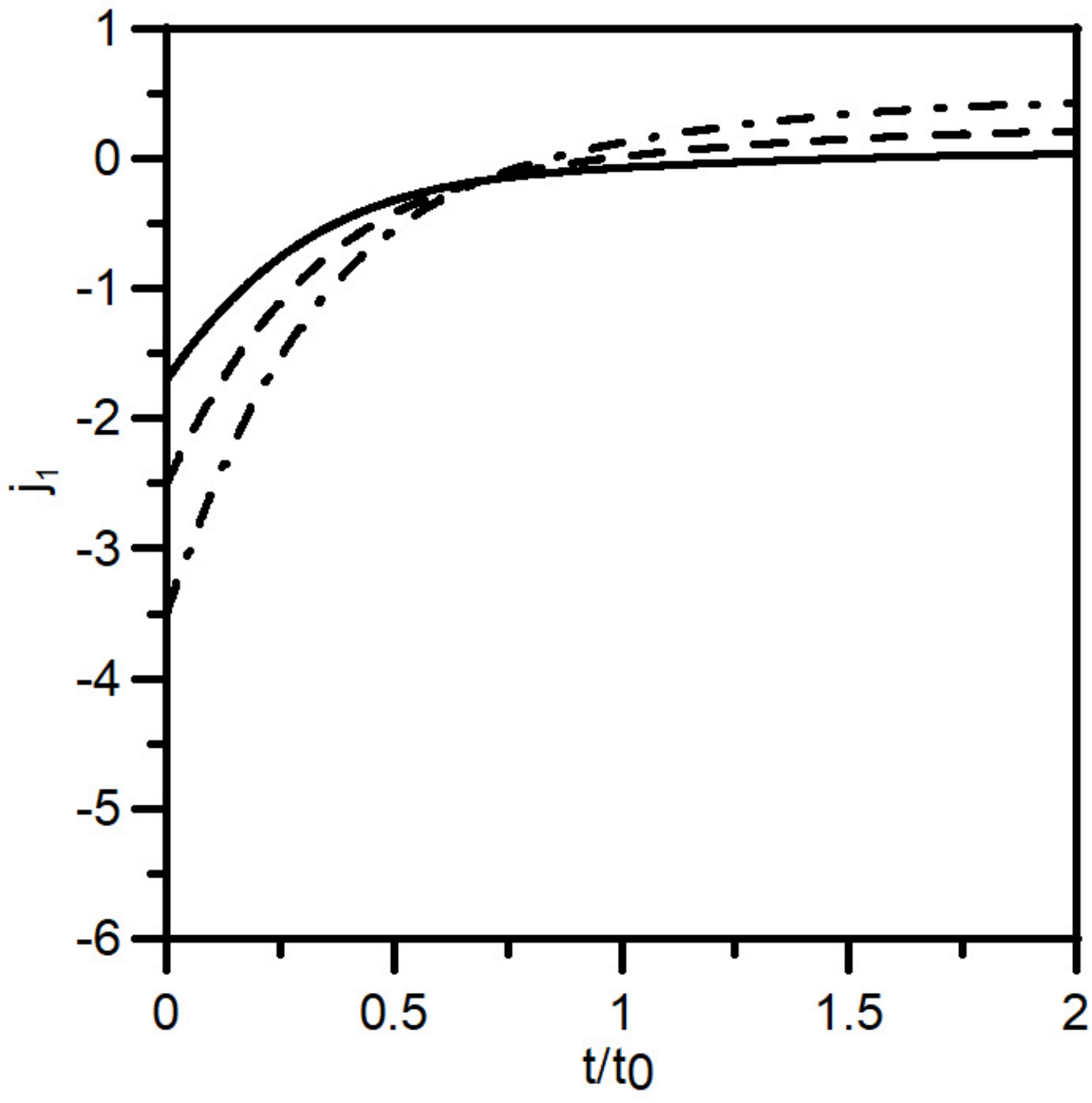}
\includegraphics[scale=0.2]{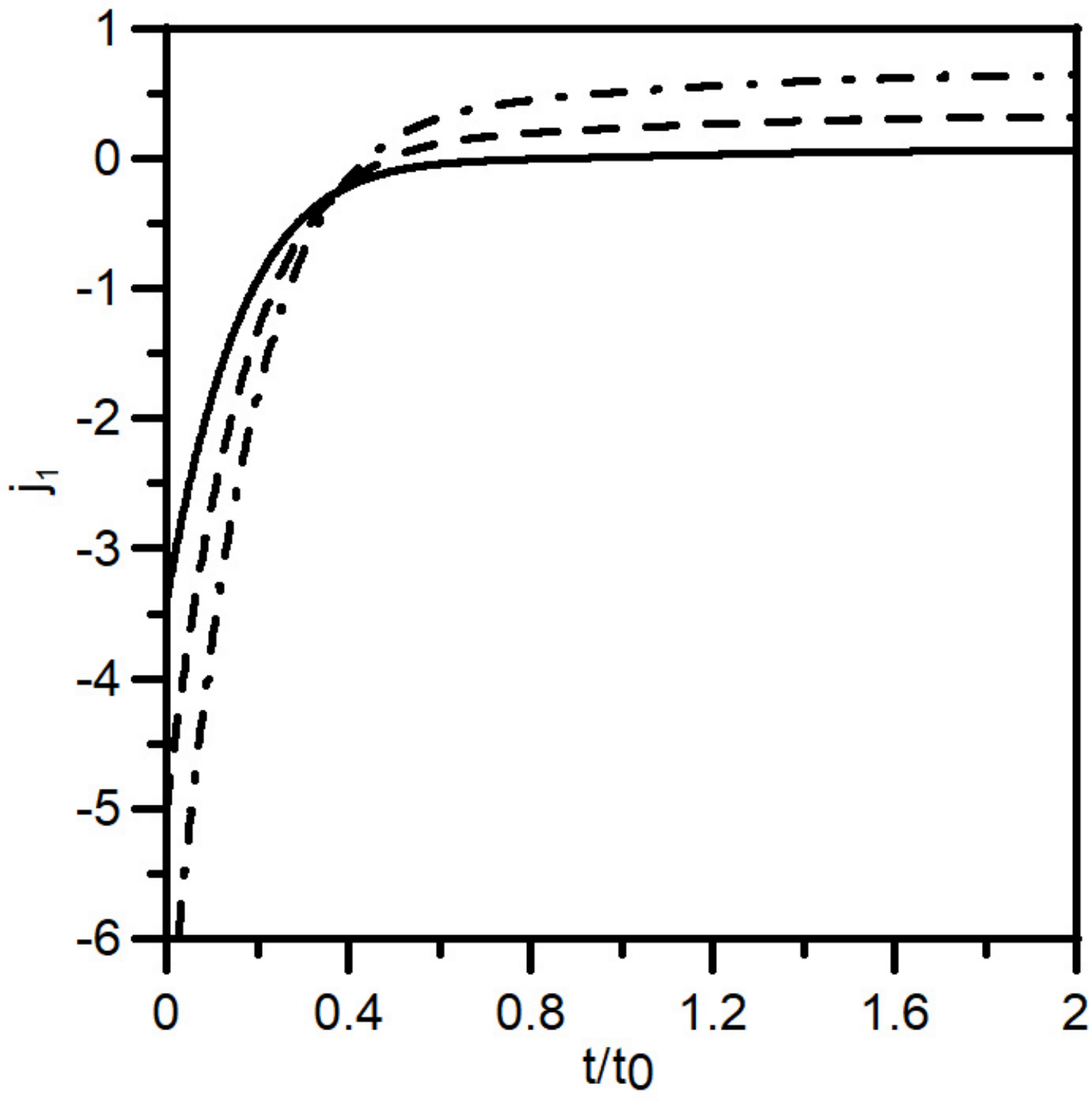}
\caption{
The time evolutions of $j_1$ for larger magnitude, $\overline{\gamma}_q = 0.15$, $0.5$, and $1$ are
shown in the left, center, and right panels, respectively.
We set $\overline{\gamma}_p = \chi = 1$ and $\overline{\beta}_2 = 1$. 
The solid, dashed, and dash-dotted lines correspond to temperatures $1/\overline{\beta}_1 = 1.2$,
$2$, and $3$, respectively, while keeping $1/\overline{\beta}_2 = 1$ fixed.
}
\label{fig:gamma_q_compa}
\end{figure}

In Figs.~\ref{fig:delta=-1} and \ref{fig:delta=-0.9}, all currents and the system energy changes
exhibit local minima.
These local minima depend on the parameter $\overline{\gamma}_q$:
they disappear for sufficiently large magnitude of $\overline{\gamma}_q$.
In Fig.~\ref{fig:gamma_q_compa}, we increase the magnitude of $\overline{\gamma}_q$ to study its
effect on the time evolution of the heat current $j_1$ with $\overline{\gamma}_p = \chi = 1$.
The panels correspond to $\overline{\gamma}_q = 0.15$ (left), $0.5$ (center), and $1$ (right). 
In each panel, the solid, dashed, and dash-dotted lines correspond to the temperatures
$1/\overline{\beta}_1 = 1.2$, $2$, and $3$, respectively, while keeping $1/\overline{\beta}_2 = 1$
fixed.
Differently from those in Figs.~\ref{fig:delta=-1} and \ref{fig:delta=-0.9} where
$\overline{\gamma}_q=0$ and $\overline{\gamma}_q=0.1$,
respectively,
all trajectories in Fig.~\ref{fig:gamma_q_compa} increase monotonically.

\begin{figure}[ht!]
\includegraphics[scale=0.2]{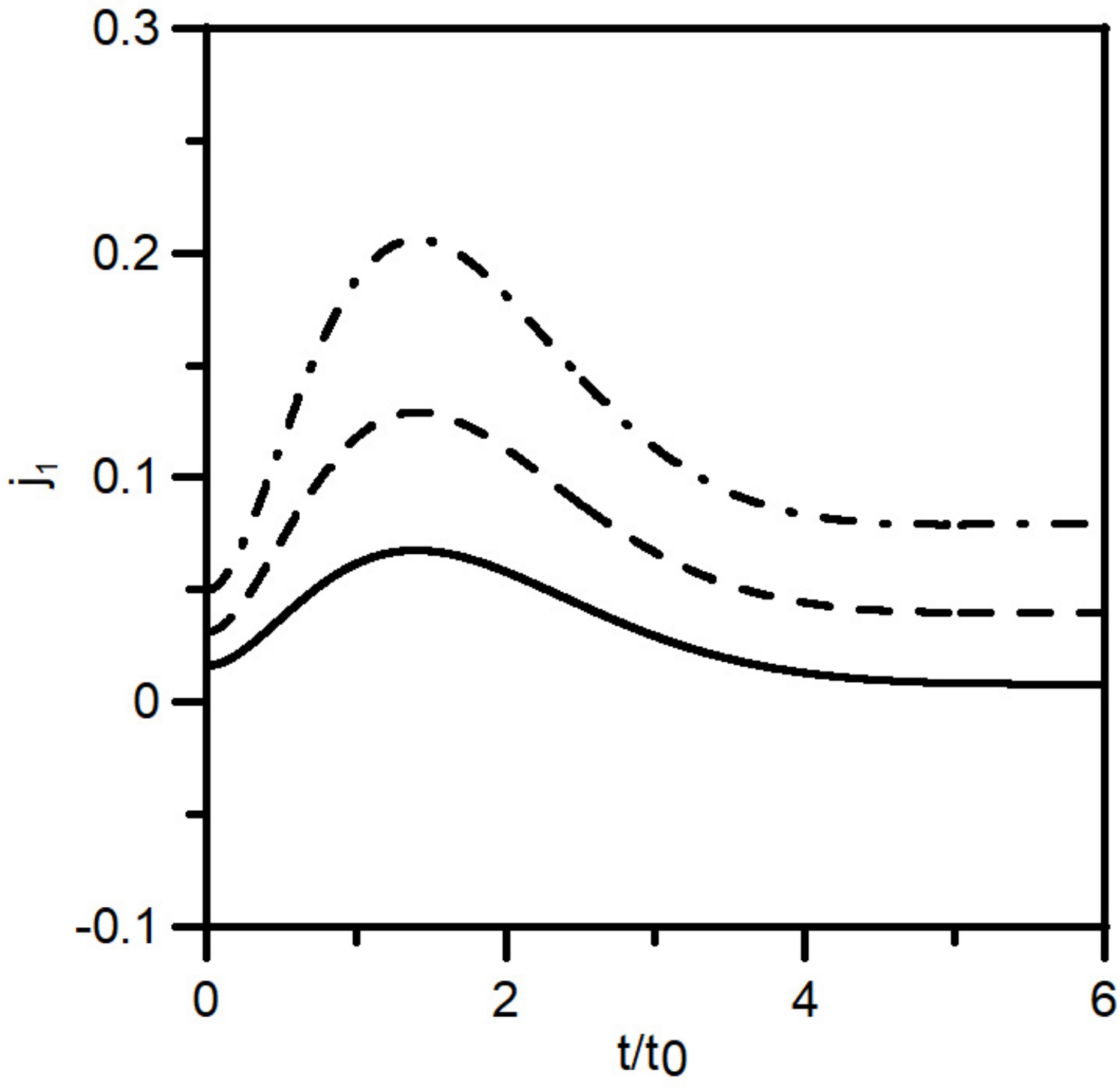}
\includegraphics[scale=0.2]{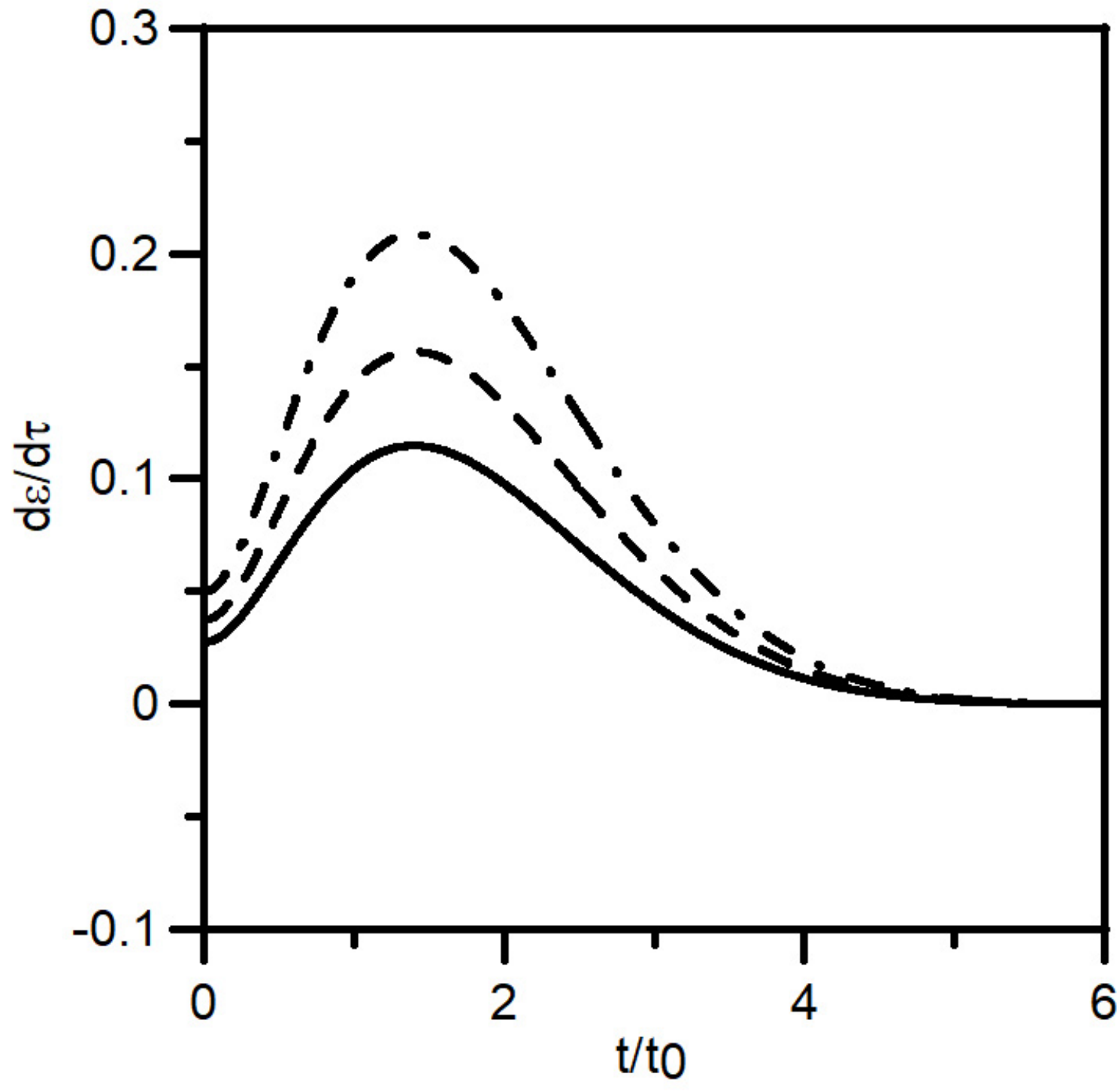}
\includegraphics[scale=0.2]{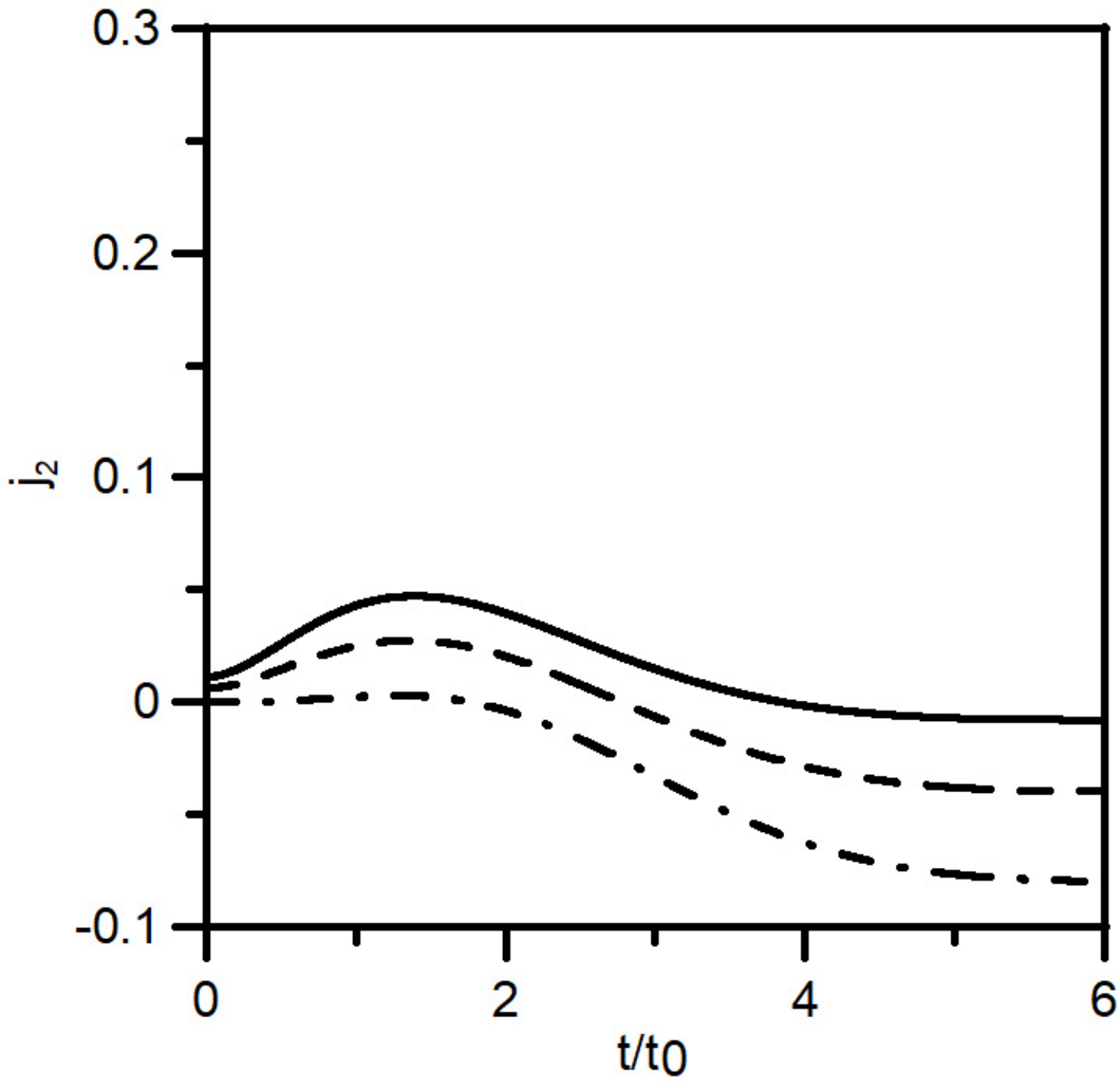}
\caption{The time evolution of the heat current $j_1$ (left panel), the rate of energy change $d\epsilon/d\tau$ (center), and the heat current $j_2$ (right panel) is plotted for $\overline{\gamma}_q = 0.1$ ($\delta = -0.9$). The parameters are set to $\chi = -0.25$ and $\overline{\beta}_2 = 1$. The solid, dashed, and dash-dotted lines correspond to the temperatures $1/\overline{\beta}_1 = 1.2$, $2$, and $3$, respectively.}
\label{fig:ki_nega}
\end{figure}

Finally, we investigate the case of a repulsive interaction between the oscillators, characterized
by a negative coupling constant ($K_\text{int} < 0$),
provided that the total potential remains stable, $-0.5 <\chi$.
Figure \ref{fig:ki_nega} shows the time evolution of the heat current $j_1$ (left panel), the
rate of energy change $d\epsilon/d\tau$ (center), and the heat current $j_2$ (right panel) for
$\overline{\gamma}_q = 0.1$ ($\delta = -0.9$).
The parameters are set to $\chi = -0.25$ and $\overline{\beta}_2 = 1$. 
The solid, dashed, and dash-dotted lines correspond to the temperatures $1/\overline{\beta}_1 =
1.2$, $2$, and $3$, respectively.
In this scenario, the introduction of the interaction at $t=0$ reduces the total potential energy of the
system. 
Consequently, the system must absorb energy from the heat baths to reach the steady state.
As shown in the figure, the initial heat currents $J_1$ and $J_2$ are both positive, indicating that heat flows from both heat baths into the system.
This behavior contrasts with the attractive case ($K_\text{int} > 0$) discussed in
Sec.~\ref{sec:numerial_gbm_posi}, yet it is entirely consistent with the principle of energy
conservation: the direction of the heat flow is governed by the energy change induced by the
interaction.

\section{Comparison of our approach with RLL Approach}

We compare our formulation with traditional methods.
The Rieder-Lebowitz-Lieb (RLL) model \cite{RLL1967} 
is considered as the historical benchmark for microscopic models of thermal conduction and still actively extended today to explore anomalous transport phenomena such as negative differential heat conductivity \cite{li2006,hu2009,Iacobucci2011,zhang2025,Krekels2026}.
This model connects Brownian thermostats to both ends of a harmonic chain. 
It demonstrated that, in the steady state, heat flow exhibits ballistic transport, meaning it remains independent of the system length. 
In this established framework, heat
current is typically defined as the mechanical power (force multiplied by velocity) provided by the
heat bath to the boundary Brownian particles.
This prescription implicitly assumes that the environment interacts only through the momentum, leaving the velocity as a well-defined, smooth quantity.
In our generalized model, however, the presence of the positional noise renders the trajectories continuous but nowhere differentiable, reflecting the inherent properties of the Wiener process.
As emphasized in the introduction, this non-standard generalization of Brownian motion is 
essential for constructing stochastic thermodynamics which is consistent with quantum thermodynamics in its classical limit \cite{KN2024,KN2024_CPTP,KN2025,Giordano}.
Consequently, the classical velocity is not well-defined, and any RLL-type definition of heat current (based on the simple product of force and velocity) would become mathematically singular.

By extending Sekimoto’s approach in stochastic thermodynamics (energetics) to incorporate both momentum and position dissipation channels \cite{KN2024}, we define heat current as the energy flux required to satisfy the first law of thermodynamics at the trajectory level. This energy-balance approach provides a rigorous and unambiguous basis for analyzing thermal transport even in systems with coordinate fluctuations, where traditional definitions fail to remain mathematically consistent.

The RLL model, considered the benchmark for microscopic models of thermal conduction, connects Brownian thermostats to both ends of a harmonic chain.
It demonstrated that, in the steady state, heat flow exhibits ballistic transport, meaning it remains independent of the system length. 
In contrast, our study aims to analytically derive the emergence of the Fourier-type linear thermal response within our generalized model of Brownian motion. 
To achieve this, we begin our investigation with the $N=2$ network model, which represents the minimal configuration.
In the standard approach by RLL \cite{RLL1967}, the local heat flow $j_{i-1 \to i}$ from site $i-1$ to $i$ is defined as the expectation value of the product of the inter-particle restoring force $\propto (\tilde{q}_{i-1} - \tilde{q}_i)$ and the velocity $d\tilde{q}_i/dt$:
\begin{equation}
j_{i-1 \to i} 
\propto \lim_{t \rightarrow \infty}\left\lceil (\tilde{q}_{i-1} - \tilde{q}_i) \dot{\tilde{q}}_i  \right\rfloor 
\propto \lim_{t \rightarrow \infty} \left\lceil \tilde{q}_{i-1} \tilde{p}_i \right\rfloor   \, ,
\label{eqn:RLL_current}
\end{equation}
In this derivation, we used $\left\lceil \tilde{q}_i \tilde{p}_i \right\rfloor  =0$ in
the asymptotic limit because
\begin{equation}
0= d \left\lceil \tilde{q}^2_i \right\rfloor = \frac{2}{m} \left\lceil \tilde{q}_i \circ d \tilde{p}_i \right\rfloor \, .
\end{equation}

Equation (\ref{eqn:RLL_current}) is, however, not applicable to our generalized model, because 
$d \tilde{q}_i \neq \tilde{p}_i dt/ m$.
The corresponding quantity in our model is 
\begin{equation}
\left\lceil \tilde{q}_i \tilde{p}_i \right\rfloor = m \gamma_q \left\lceil \tilde{q}_i \frac{\partial H}{\partial \tilde{q}_i} \right\rfloor - \frac{m \gamma_q}{\beta_i} \, .
\end{equation}
This indicates that in our generalized model ($\gamma_q \neq 0$), this correlation persists as a non-zero value, suggesting that the traditional RLL-style definition of heat flow should be modified.
The corresponding modification of the definition of heat in stochastic thermodynamics,
considered as the work due to the action of the bath, 
is straightforward as shown in Ref.~\cite{KN2024}.

\section{Conclusion}
\label{sec:conclusion}

In this paper, we systematically investigated the heat conduction properties of a coupled harmonic oscillator network described by a generalized model of Brownian motion. 
Motivated by the consistency requirements with quantum open systems, this framework 
incorporates fluctuation and dissipation effects into both the momentum and position equations, thereby satisfying the fluctuation-dissipation theorem. 
Consequently, the standard proportional relation between velocity and momentum is no longer preserved. While this modification may appear unconventional, the deviation from a simple linear relation between momentum and velocity is a natural feature in the presence of velocity-dependent forces, such as the Lorentz force in electromagnetism. 
Crucially, incorporating these specific fluctuation and dissipation effects into the position coordinate is known to be strictly necessary to correctly reproduce the classical limit of the Gorini-Kossakowski-Sudarshan-Lindblad (GKSL) equation describing open quantum systems \cite{KN2024,KN2024_CPTP,KN2025,Giordano}. 
As a direct mathematical consequence of this generalized structure, the noise term acting on the position coordinate yields trajectories that are continuous but nowhere differentiable. 
This behavior uniquely reflects the inherent properties of the Wiener process, contrasting with the standard model where the position trajectory remains smooth.

First, we analytically demonstrated that in the steady state of the coupled harmonic oscillator network, the heat current is proportional to the temperature difference between the two heat baths. This confirms that Fourier's law (linear thermal response) is successfully reproduced not only in standard Brownian motion but also in generalized Brownian motion. 
Consequently, the generalized Brownian motion, which was introduced to ensure consistency with quantum thermodynamics, serves as a fundamentally appropriate model for constructing stochastic thermodynamics, even in the context of mesoscopic heat conduction phenomena.

Furthermore, our numerical analysis of the transient dynamics revealed striking non-equilibrium behaviors. 
Most notably, the presence of position dissipation ($\gamma_q \neq 0$) induces an instantaneous
non-zero heat current
at the moment the interaction between the harmonic oscillators is suddenly turned on.
%
We demonstrated that the direction of this initial transient heat flow is strictly governed by the principle of energy conservation and depends on the sign of the interaction potential.
An attractive interaction triggers an immediate energy dissipation into the baths, whereas a repulsive interaction forces the system to absorb energy from the environment
during the relaxation toward a new steady state.
However, it should be emphasized that this
initial instantaneous heat current is a
direct consequence of the idealized assumption that the interaction is introduced abruptly as a step function at $t=0$ to two independent thermally equilibrated harmonic oscillators.
If the interaction were instead turned on gradually over a finite time interval, as would be
expected in a realistic experimental setup, this instantaneous jump would naturally disappear.
Further systematic studies on the system dynamics under a time-dependent coupling $K_{\text{int}}(t)$ will clarify more realistic non-equilibrium transient behaviors.

Our analytical treatment further captured the temperature discontinuity at the system-bath
interface, providing a microscopic manifestation of thermal boundary resistance, analogous to
Kapitza resistance \cite{Lepri2003,Swartz1989,Giri2020}. Specifically, by evaluating the effective
temperatures of the interacting oscillators assuming the equipartition theorem, we found that the
internal temperature difference of the system is strictly smaller than the temperature difference
between the external heat baths. This incomplete thermalization demonstrates how finite boundary
coupling induces a temperature jump, indicating that the model is capable of capturing realistic
interfacial transport phenomena.

These findings demonstrate that the generalized framework of Brownian motion provides a consistent and phenomenologically rich description of non-equilibrium steady processes. 
Furthermore, the quantum open dynamics obtained by quantizing this classical model has already been studied; it is known that, with an appropriate choice of parameters, the resulting dynamics satisfies the CPTP condition, thereby strictly fulfilling the fundamental requirements of quantum mechanics \cite{KN2024,KN2024_CPTP,KN2025,Giordano}. 
By further utilizing our model, it will be possible to systematically investigate how properties
specific to quantum systems are reflected in heat conduction phenomena.
See, for example, recent papers \cite{Weiderpass2020,Babakan2026a} and references therein.

More specifically, the quantized model and its classical limit allow us to place
the generalized model of Brownian motion within the ongoing discussion on global versus
local quantum master equations \cite{levy,volovich,cattaneo}.
The distinction between these approaches lies in whether the dissipative dynamics is
constructed in the eigenbasis of the full interacting system Hamiltonian (global)
or in the basis of uncoupled subsystems prior to diagonalization (local).
In our case, the quantization procedure naturally leads to a global-type master equation,
since dissipation and system dynamics are consistently derived from the same interacting
Hamiltonian, rather than being introduced independently at the subsystem level.
In this sense, the global character emerges from the unified treatment
of system–bath interactions, an intrinsic feature of our generalized framework
\cite{KN2024,KN2024_CPTP,KN2025}.
While the local–global debate has been extensively explored
for similar systems within standard GKSL descriptions \cite{gonzalez},
our formulation provides an alternative perspective grounded in stochastic thermodynamics.
From a quantum-thermodynamic standpoint, this framework serves as a useful benchmark: it
demonstrates that a description consistent with the CPTP dynamics at the quantum level can simultaneously reproduce the expected Fourier-type linear heat
conduction in the classical limit. 
Heat transport described by the GKSL equations in related
systems has been investigated in Refs.~\cite{assadian,nicacio7,gonzalez},
further supporting this connection.

Another significant direction for future research is the extension of this framework to
one-dimensional lattices ($N \gg 2$).
As extensively discussed in the context of low-dimensional statistical mechanics, standard momentum-conserving systems often exhibit anomalous heat conduction, where the thermal conductivity diverges with the system size rather than converging to a finite value \cite{RLL1967,Lepri2003,Dhar2008}. 
This breakdown of Fourier's law is typically attributed to the unhindered ballistic transport of phonons in integrable systems.
However, our generalized model introduces fluctuation and dissipation explicitly into the position coordinates ($\gamma_q$). 
This unique feature may act as an intrinsic scattering mechanism that effectively breaks the ballistic propagation of energy carriers, even in harmonic chains. 
It remains an open and crucial question whether the positional dissipation 
$\gamma_q$ is sufficient to suppress anomalous transport and restore normal diffusive behavior (finite thermal conductivity) in the thermodynamic limit remains an open and crucial question.

\begin{acknowledgments}
T.K. acknowledges the financial support by CNPq (No.\ 304504/2024-6). 
A part of this work has been done under the project INCT-Nuclear Physics and Applications (No.\
464898/2014-5;408419/2024-5.).
F.N. is a member of the Brazilian National Institute of
Science and Technology in Quantum Devices (INCT-DQ)
(CNPq, Grant No. 408783/2024-9).
\end{acknowledgments}


\end{document}